\documentclass[twocolumn,5p]{elsarticle}
\pdfoutput=1
\usepackage{hyperref}
\usepackage{tikz}

\usepackage{comment}
\usepackage{framed,graphicx}
\usepackage{multirow}
\usepackage{rotating}
\usepackage{amsmath}
\usepackage{bigstrut}
\usepackage{color}
\usepackage{graphics}
\usepackage{eqparbox}
\usepackage{graphics}
\usepackage{colortbl}
\usepackage{mathptmx} \usepackage[scaled=.90]{helvet} \usepackage{courier}
\usepackage{balance}
\usepackage{picture}
\usepackage{algorithm}
\usepackage{algorithmicx}
\usepackage{algpseudocode}
\usepackage[export]{adjustbox}
\renewcommand{\footnotesize}{\scriptsize}
\definecolor{lightgray}{gray}{0.8}
\definecolor{darkgray}{gray}{0.6}
\definecolor{lavenderpink}{rgb}{0.98, 0.68, 0.82}
\definecolor{celadon}{rgb}{0.67, 0.88, 0.69}

\newcommand{\quart}[4]{\begin{picture}(80,4)%1
	{\color{black}\put(#3,2){\circle*{4}}\put(#1,2){\line(1,0){#2}}}\end{picture}}

\definecolor{Gray}{gray}{0.95}
\definecolor{LightGray}{gray}{0.975}

\newcommand{\rahul}[1]{{\color{steel}{#1}}}

\newcommand{\bi}{\begin{itemize}[leftmargin=0.4cm]}
	\newcommand{\ei}{\end{itemize}}
\newcommand{\be}{\begin{enumerate}}
	\newcommand{\ee}{\end{enumerate}}
\newcommand{\tion}[1]{\S\ref{sect:#1}}
\newcommand{\fig}[1]{Figure~\ref{fig:#1}}

\newcommand{\eq}[1]{Equation~\ref{eq:#1}}

\usepackage[shortlabels]{enumitem}
\usepackage{url}

\definecolor{steel}{rgb}{0,0,0}
\definecolor{steel2}{rgb}{.11, .11, .7}
\definecolor{Gray}{rgb}{0.88,1,1}
\definecolor{Gray}{gray}{0.85}
\usepackage[framed]{ntheorem}
\usepackage{framed}
\usepackage{tikz}
\usetikzlibrary{shadows}
\theoremclass{Lesson}
\theoremstyle{break}

\tikzstyle{thmbox} = [rectangle, rounded corners, draw=black,
fill=Gray!40,  drop shadow={fill=black, opacity=1}]

\newshadedtheorem{lesson}{Result}

\begin{document}
	
	\begin{frontmatter}
		
		\title{Less is More: Minimizing  Code Reorganization using XTREE}
		
		\author{Rahul Krishna\corref{cor1}\textsuperscript{a,}}
		\ead{i.m.ralk@gmail.com}
		\author{Tim Menzies\corref{cor1}\textsuperscript{a,}}
		\ead{tim.menzies@gmail.com}
		\author{Lucas Layman\textsuperscript{b}}
		\ead{ llayman@cese.fraunhofer.org }
		\cortext[cor1]{Corresponding author: Tel:+1-919-396-4143(Rahul)}
		\address{\textsuperscript{a}Department of Computer Science, North 
		Carolina State University, Raleigh, NC, USA\\
			\textsuperscript{b}Fraunhofer CESE, College Park, USA}
		\pagenumbering{arabic}
		
		\begin{abstract}
			{\bf Context: }
			Developers use bad code smells to guide code reorganization.
			Yet developers, textbooks, tools, and researchers disagree on 
			which bad smells are important. How can we offer reliable
                        advice to developers about which bad smells to fix?
			
			\noindent
			{\bf Objective:} To evaluate the likelihood that a code 
			reorganization to address bad code smells will yield improvement in 
			the defect-proneness of the code.
			
			\noindent
			{\bf Method: } We introduce XTREE, a framework that analyzes a 
			historical log of defects seen previously in the code and generates 
			a set of useful code changes.
			Any bad smell that requires changes outside of that set can be 
			deprioritized (since there is no historical evidence that the bad 
			smell causes any problems).
			
			\noindent
			{\bf Evaluation: } We evaluate XTREE's recommendations for bad 
			smell improvement against recommendations from previous work 
			(Shatnawi, Alves, and Borges) using multiple data sets of code 
			metrics and defect counts.
			
			\noindent
			{\bf Results: }Code modules that are changed in response to XTREE's 
			recommendations contain significantly fewer defects than 
			recommendations from previous studies. Further, XTREE endorses 
			changes to very few code metrics, so XTREE requires programmers to 
			do less work. Further, XTREE's recommendations are more 
            responsive to the particulars of different data sets. Finally 
            XTREE's recommendations may be generalized to identify the most 
            crucial factors affecting multiple datasets (see the last figure in 
            paper).
                        		
			\noindent
			{\bf Conclusion: }
			Before undertaking a code reorganization based on a bad smell 
			report, use a framework like XTREE to check and ignore any such 
			operations that  are useless; i.e. ones which lack evidence in the 
			historical record  that it is useful to make that change. Note 
			that this use case applies to both manual code reorganizations 
			proposed by developers as well as those conducted by automatic 
			methods.

		\end{abstract}
	\end{frontmatter}
	\pagenumbering{arabic} %XXX delete before submission
	
	\vspace{1mm}

	\section{Introduction}
	\label{sect:intro}
	According to   Fowler~\cite{fowler99}, bad smells (a.k.a. code smells)
	are ``a surface indication that usually corresponds to a deeper problem''.
	Fowler  recommends   removing   code smells   by
	\begin{quote}
		``$\ldots$ applying a series of small behavior-preserving 
		transformations, 
		each
		of which seem `too small to be worth doing'.
		The  effect of   these refactoring transformations is quite 
		significant. By doing them in small steps you reduce the risk
		of introducing errors''.
	\end{quote}

	While the original concept of bad smells was largely subjective, 
	researchers including Marinescu~\cite{Lanza2006} and 
	Munro~\cite{munro2005product} provide definitions of ``bad smells'' in 
	terms of static code attributes such as size, complexity, coupling, and 
	other metrics. Consequently, code smells are captured by popular static 
	analysis tools, like PMD\footnote{https://github.com/pmd/pmd}, 
	CheckStyle\footnote{http://checkstyle.sourceforge.net/}, 
	FindBugs\footnote{http://findbugs.sourceforge.net/}, and 
	SonarQube\footnote{http://www.sonarqube.org/}.

        We refer to the process of removing bad smells as \textit{code 
	reorganization}. Code reorganization is an amalgum of perfective and 
	preventive maintenance~\cite{iso14764}. In contrast to refactoring, code 
	reorganization is not guaranteed to preserve behavior.
	Fowler~\cite{fowler99} and other influential software 
	practitioners~\cite{mcconnell2004code,horror} recommend refactoring and 
	code reorganization to remove bad smells. Studies suggest a relationship 
	between code smells and poor maintainability or defect 
	proneness~\cite{yamashita2013code,yama13,zazworka2011investigating}, though 
	these findings are not always consistent~\cite{olbrich2010all}.
	
	The premise of this paper is that not every bad smell needs to be fixed.
        For example, 
	the origins of this paper was a meeting with a  Washington-based software company where managers
        shared one of 
	their challenges: their  releases were delayed by 
	developers spending  much time removing bad smells within their code. There is much evidence
        that this is a common problem. Kim 
	et al.~\cite{kim2012field} surveyed developers at Microsoft and found that 
	code reorganizations incur significant cost and risks. Researchers are 
	actively attempting to demonstrate the actual costs and benefits of fixing 
	bad 
	smells~\cite{nugroho2011empirical,zazworka2011prioritizing,zazworka2013case},
	 though more case studies are needed.
	
	 In this 
	paper, we focus on the challenge of recommending code reorganizations that 
	result in perceivable benefits (such as reduced defect proneness) and avoid 
	those reorganizations which have no demonstrable benefit and thus waste 
	effort.
	 To this end, this
	paper evaluates XTREE~\cite{krishna2015actionable}, a framework to evaluate 
	whether a code reorganization is likely to have a perceivable benefit in 
	terms of defect-proneness. We focus on bad smells indicated by code 
	metrics 
	such as size and complexity as captured in popular tools such as SonarQube 
	and Klockwork\footnote{http://www.klocwork.com/}.
	XTREE examines the historical record of code metrics for a project. If 
	there is no evidence that changing code metric ``X'' is useful (e.g., 
	lowering ``X'' reduces defect-proneness), then developers should be 
	discouraged from wasting effort on that change.
	
	Our method uses two oracles: a {\em primary
		change oracle} and a {\em secondary verification oracle}.
	By combining these two oracles,
	we can identify and validate useful
	code reorganizations.
	
	We use
	the XTREE cluster delta algorithm as the {\em primary change  oracle}.
	XTREE
	explores the historical record of a project to find clusters of modules 
	(e.g., files or binaries).
	It then proposes a ``minimal'' set of changes $\Delta$ that can move a 
	software module $M$ from a defective cluster $C_0$ to another with fewer 
	defects $C_1$ (so $\Delta$
	is some subset of $C_1 - C_0$). % To define ``minimal'', XTREE uses an 
	%entropy measure.
	
	The {\em secondary verification oracle} checks if the primary oracle is 
	proposing
	sensible changes. We create the verification oracle using Random 
	Forest~\cite{Breiman2001} augmented with SMOTE (synthetic  minority 
	over-sampling technique~\cite{chawla2002smote}).
	In our framework,  learning
	the secondary oracle is   a {\em separate} task from that of learning the 
	primary
	oracle. This  ensures that the verification oracle offers an independent
	opinion on the value of the proposed changes.
	
	An advantage to the XTREE cluster delta approach is that it avoids the
	{\em conjunctive fallacy}.
	A common heuristic in the bad smell 
	literature~\cite{erni96,bender99,Shatnawi10,Alves2010,hermans15} is: for 
	all static code measures that exceed some threshold, make changes such that 
	the thresholds are no longer exceeded. That is:
	\begin{equation}\label{eq:df}
		\begin{array}{rl}
			\mathit{bad}    & = \left(a_1 > t_1 \right) \bigvee \left(a_2 > 
			t_2\right) \bigvee    ...                             \\
			\mathit{better} & = \neg\;\mathit{bad} = \left(a_1 \le t_1 \right) 
			\bigwedge \left(a_2 \le t_2\right)  \bigwedge  ... 
		\end{array}
	\end{equation}
	We say that the above definition of ``better'' is a conjunctive fallacy
	since it assumes that the best way to improve code is to decreases multiple 
	code attribute measures to below $t_i$ in order
	to remove the ``bad'' smells. In reality, the associations between static 
	code measures are more intricate, for example, {\em decreasing}  $a_i$ may 
	necessitate {\em increasing} $a_j$.
	It is easy to see why this is so.
	For example, Fowler recommends that the Large Class smell be addressed by 
	the Extract Class or Extract Subclass refactoring~\cite{fowler99}. If we 
	pull code out of a function since that function has grown too
	large, that functionality has to go somewhere else. Thus, when 
	\textit{decreasing} lines of code in one module, we may \textit{increase} 
	its coupling to another module. XTREE identifies such associations between 
	metrics, thus avoiding the conjunctive fallacy.
	
	\subsection{Research Questions}
	\label{sect:rqs}
	This paper  claims that (a)~XTREE is more useful than \eq{df} to identify 
	code reorganizations, and (b)~XTREE recommends a small subset of static 
	code measures to change and thus can be used to identify superfluous code 
	reorganizations (i.e., reorganizations based on omitted
	static code measures). 
	
	\rahul{To evaluate these claims, we compared the 
	performance of XTREE with three other methodologies/tools for recommending 
	code reorganizations: (1) VARL based thresholds~\cite{Shatnawi10}; (2) 
	Statistical threshold generation~\cite{Alves2010}; (3) CD, cluster based 
	framework~\cite{me12c} according to the following research questions:}

	{\bf  RQ1: Effectiveness}: According to the verification oracle, which of 
	the frameworks introduced above is most accurate in recommending code 
	reorganizations that 
	result in reduced numbers of defective modules?

	To answer this question, we used data from five OO Java projects
	(Ivy, Lucene, Ant, Poi, Jedit). It was found that:
	\begin{lesson}
		XTREE is the most accurate oracle on how to change code modules in 
		order to reduce the number of defects.
	\end{lesson}
	
	{\bf RQ2: Succinctness}: Our goal is to critique and, possibly,
	ignore irrelevant recommendations for removing bad smells.  If the 
	recommended changes are 
	minimal (i.e. affect fewest attributes) then those changes
	will be easiest to apply and monitor. So, which framework recommends 
	changes to the fewest code attributes?
	\begin{lesson}
		Of all the code change oracles studied, XTREE recommends changes to the 
		fewest number of static code measures.
	\end{lesson}
	
	{\bf RQ3: Stopping}: Our goal is to discourage code reorganization based on 
	changes that lack
	historical evidence of being effective. So, how effective is XTREE at 
	identifying what {\em not} to change?
	\begin{lesson}
		In  any  project,  XTREE's  recommended  changes  to 1--4
		of the  static code attributes.  Any bad smell defined in terms of the 
		remaining 19 to 16 code attributes (i.e. most of them)
		would hence be deprecated.
	\end{lesson}
	
	{\bf RQ4: Stability}: Across different projects, how consistent are the 
	changes recommended by our best change oracle?
	\begin{lesson}
		The direction of change recommended by XTREE (e.g., to LOWER lines of 
		code while RAISING coupling) is stable across repeated runs of change 
		oracle.
	\end{lesson}
	
	{\bf RQ5: Conjunctive Fallacy}:
	Is it always  useful  to apply \eq{df}; i.e. make code better by   
	reducing 
	the values of  multiple code attributes? We find that:
	\begin{lesson}
		XTREE usually recommends reducing lines of code (size of the modules).
		That said,  XTREE often recommends {\em increasing} the values of other 
		static code attributes.
	\end{lesson}
	Note that {\bf RQ3, RQ4, RQ5}
	confirms the intuitions
	of the project managers that prompted this investigation. We find evidence 
	that:
	\bi
	\item Code reorganizations that decrease multiple measures may not yield 
	improvement. In fact,  programmers may need to {\em decrease} some 
	measures 
	while {\em increasing} others.
	\item In the studied projects, XTREE recommends changes to approximately 
	20\% of the code measures, and thus reorganizations based on the remaining 
	80\% are unlikely to provide benefit.
	\ei

	In terms of concrete recommendations for
		practitioners, we say {\em look before you leap}:
		\bi
		\item Before doing  code reorganization based on a bad smell 
		report$\ldots$ 
		\item $\ldots$ check and discourage any code reorganization   for which 
		there 
		is no proof
		in the historical record that the change improves the code.
		\ei
		Aside: this recommendation   applies to both manual code 
		reorganizations proposed by developers
		as well as the code reorganizations conducted by automatic 
		methods~\cite{mkaouer2015many}. That is, XTREE could optimize automatic
		code reorganization by discouraging reorganizations for useless goals.

	Beside this introduction, the rest of this paper is formatted as follows. 
	\tion{prior0} relates the current work to the prior literature on bad 
	smells and code reorganization efforts. In \tion{dataset}, we 
	discuss the choice 
	of our datasets and briefly describe it's structure. \tion{bst} details the 
	frequently used techniques on leaning bad smell thresholds to assist 
	reorganization. Specifically, 
	\tion{xtreextree} and \tion{xtreextree2} describe our preferred method for 
	performing code reorganizations. Our experimental setup and evaluation 
	strategies are presented in \tion{setup}. The results and corresponding 
	descriptions are available in \tion{results}. The future work and the 
	reliability of our finding are available in \tion{future} and \tion{valid} 
	respectively. Finally, the conclusions are presented in \tion{conclusions}.

	\section{Relationship to Prior Work }
	\label{sect:prior0}
	{\color{steel}	
	\subsection{Prioritizing Reorganization Efforts}
	\label{sect:prior}

		There have been several efforts to prioritize refactoring efforts. 
		These attempts address code smells in particular and have garnered more 
		attention over the past few years. Ouni et al.~\cite{ouni3,ouni1,ouni2} 
		use search 
		based software engineering to suggest refactoring solutions. They 
		recommend the use of four 
		factors: (1) priority; (2) severity; (3) risk; and (4) importance, all 
		of which are determined by developers. Both the developer's 
		recommendations and the aforementioned factors can and do vary over 
		time especially as classes are modified. Additionally, they also vary 
		with projects. 
		
		A similar direction was taken by Vidal et al.~\cite{vidal14}. They 
		presented a semi-automated approach for prioritizing code smells. They 
		then recommend a suitable refactoring based on a developer survey. The 
		determination of severity is based primarily on three criteria: (1) the 
		stability of the component in which the smell was found; (2) the 
		subjective assessment that the developer makes of each kind of smell 
		using an ordinal scale; and (3) the related modifiability scenarios.
		
		The standard approach is to develop and evaluate these findings by 
		interviewing human developers. We, however, dissuade practitioners 
		from taking this approach for reasons discussed in depth in 
		\tion{prelim}.
		
		In a more recent study, a slightly different approach was proposed by 
		Vidal et al.~\cite{font1}. In their work, static code metrics are used 
		to detect the presence of code smells. This is followed by a ranking 
		scheme to measure the severity of code smells. This was a fully 
		automated approach. However, the authors fail to report the accuracy 
		of 
		detection of code smells. This issue is further compounded by the use 
		of mean and standard deviation of metrics to detect the presence of 
		code smells which has been criticized by several  
		researchers~\cite{Shatnawi10,Alves2010}. Those authors do acknowledge 
		that the prioritization of detection results obtained using their 
		"intensity" measure, at the time of publication, lacked comprehensive 
		experimental validation.

		\subsection{Preliminary Report on Code Reorganization}
		There is a distinction between our work and all methods listed above. 
		The code smell prioritization efforts assist developers in choosing 
		\textit{which} refactoring operation to undertake first. Instead of 
		prioritizing refactoring, our work places more focus on assisting 
		developers by recommending useful code changes which in turn helps 
		deprioritize certain code reorganization efforts. Our preferred 
		framework 
		to 
		achieve this is XTREE.}
	
	XTREE was first introduced as a four page preliminary 
	report\footnote{https://goo.gl/2In3Lr} which was presented 
	previously~\cite{krishna2015actionable}. That short report offered case 
	studies on only two  of
	the five data sets studied here. We greatly expand on this prior work by:
	\bi
	\item
	Evaluating XTREE's recommendations against recommendations from frameworks 
	developed using three other methods proposed by other researchers exploring 
	bad 
	smells.
	\item
	Evaluating if XTREE's and other methods' recommended changes were sensible 
	using the {\em secondary verification oracle}
	\ei
	Note that only sections \tion{cdcd}, \tion{xtreextree}, and two-fifths of the results 
	in \fig{jur2} 
	contain material
	found in prior papers.

	\subsection{Why Not Just Ask Developers to Rank Bad 
	Smells?}\label{sect:prelim}
	
	Why build tools like XTREE to  critique proposed developer actions?
	Much research endorses code smells as a guide for
	code improvement (e.g., code reorganization or preventative maintenance). 
	A 
	recent literature review by Tufano et al.~\cite{Tufano2015}
	lists dozens of papers on smell detection and repair tools.
	Yet
	other papers cast doubt on the value of bad smells
	as triggers for code 
	improvement~\cite{Mantyla2004,Yamashita2013,Sjoberg2013}.
	
	If the SE literature is contradictory, why not ignore it and use domain 
	experts (software engineers) to decide
	what bad smells to fix? We do not recommend this since developer {\em 
	cognitive biases} can mislead them to
	assert that some things are important and relevant when they are not.
	According to Passos et al.~\cite{passos11},  developers often  assume that 
	the lessons they learn from a few past projects are general to all their 
	future projects. They comment, ``past experiences were taken into account 
	without much consideration for their context''~\cite{passos11}.  
	J{\o}rgensen \& Gruschke~\cite{jorgensen09} offer a similar warning. They 
	report that the supposed software engineering ``gurus'' rarely use lessons 
	from past projects to improve their future reasoning and that such poor 
	past advice can be detrimental to new projects.~\cite{jorgensen09}.

        Other 
	studies have shown some widely-held views are   now questionable given new 
	evidence.
	Devanbu et al. examined responses from 564 Microsoft software developers 
	from around
	the world. They comment programmer beliefs can vary with each project, but do not necessarily
	correspond with actual evidence in that project~\cite{prem16}.
	\def\checkmark{\tikz\fill[scale=0.4](0,.35) -- (.25,0) -- (1,.7) -- (.25,.15) -- cycle;} 
\begin{figure}[!t] 
\scriptsize
\centering
\begin{tabular}{r|c|c|c|c}

\begin{turn}{75}Fowler'99~\cite{fowler99} and~\cite{Kerievsky2005}\end{turn} &\begin{turn}{75} Lanza'06~\cite{Lanza2006}\end{turn} & \begin{turn}{75}SonarQube~\cite{sq15} \end{turn} &  \begin{turn}{75}Yamashita'13\cite{Yamashita2013} \end{turn}& \begin{turn}{75} Developer Survey 2015\end{turn}\\\hline
  Alt. Classes with Diff. Interfaces & & & & \\
  Combinatorial Explosion~\cite{Kerievsky2005} & & &  & \\
  Comments & & & 11 & VL\\
  Conditional Complexity~\cite{Kerievsky2005} & & & 14  & ?\\
  Data Class & \checkmark & &  &\\
  Data Clumps &  &  &  &\\
  Divergent Change & & &  & \\
  Duplicated Code & \checkmark & \checkmark & 1  & VH\\
  Feature Envy & \checkmark & & 8  &\\
  
  Inappropriate Intimacy & & \checkmark &  & L\\
  Indecent Exposure~\cite{Kerievsky2005} & & &  & ?\\
  Incomplete Library Class & & &  &\\
  Large Class & \checkmark & \checkmark & 4  & VH\\
  Lazy Class/Freeloader & & \checkmark & 7  &\\
  Long Method & \checkmark& \checkmark & 2  & VH\\
  Long Parameter List &  & \checkmark & 9  & L \\
  
  Message Chains & & &  & H\\
  Middle Man & &  &  &\\
  Oddball Solution~\cite{Kerievsky2005} & & &  & \\
  Parallel Inheritance Hierarchies & & &  &\\
  Primitive Obsession &  & &  &\\
  
  Refused Bequest & \checkmark & \checkmark &  & \\ 
  Shotgun Surgery & \checkmark& &  & \\
  Solution Sprawl~\cite{Kerievsky2005} & & &  &\\
  Speculative Generality & & &  & L\\
  Switch Statements &  & &  & L\\
  Temporary Field & & \checkmark &  & ?\\
  \end{tabular}
%   \multicolumn{5}{l}{} \\
%   \multicolumn{5}{l}{\textsuperscript{\textdagger} + Exact rule, - Possibly related rule} \\
%     \multicolumn{5}{l}{\textsuperscript{\textdaggerdbl} + is included in list} \\
%     \multicolumn{5}{l}{\textsuperscript{\textexclamdown} VH: Very High, H: High, L: Low, VL: Very Low, ?: Still investigating} \\
%   \multicolumn{5}{l}{\textsuperscript{*} Computation is based on more than simple metrics.} \\

\caption{Bad   smells from different sources.  Check marks (\protect\checkmark) denote   a bad smell was mentioned.
Numbers or symbolic labels (e.g. "VH") denote  a priorization comment (and
``?'' indicates lack of consensus). Empty cells
denote some bad smell listed in column one that was not found relevant
in other studies.
Note: there are many blank cells.}
\label{fig:smells}
\end{figure}
	If the above remarks hold true for bad smells, then we would expect
	to see much disagreement on which bad smells are important and relevant
	to  a particular project.

        This is indeed the case.
	The first column of \fig{smells}
	lists  commonly mentioned bad smells and comes from Fowler's 1999 
	text~\cite{fowler99} and a subsequent 2005 text by Kerievsky that is widely 
	cited~\cite{Kerievsky2005}.
	The other
	columns show data from other studies on which bad smells matter most.
	The columns marked as Lanza'06 and Yamashita'13 are from peer reviewed 
	literature. The column marked SonarQube is a popular open source
	code assessment tool that includes detectors for six of the bad smells
	in column one.
	The {\em developer survey} (in the right-hand-side column) shows the 
	results of an hour-long whiteboard session with a group of 12 developers 
	from a Washington
	D.C. web tools development company. Participants
	worked in a round robin manner to rank the bad smells they thought were
	important (and any disagreements were discussed with the whole group).
	Amongst the group, there was  some
	consensus on  the priority of which bad smells to fix
	(see the annotations VH=very high,
	H=high, L=low, VL=very low, and ``?''= no consensus).

	\begin{figure*}[t!]
	\renewcommand{\baselinestretch}{0.8}\begin{center}
		{\scriptsize
			\begin{tabular}{c|l|p{4.7in}}
				amc & average method complexity & e.g. number of JAVA byte codes\\\hline
				avg\, cc & average McCabe & average McCabe's cyclomatic complexity seen
				in class\\\hline
				ca & afferent couplings & how many other classes use the specific
				class. \\\hline
				class. \\\hline
				cam & cohesion amongst classes & summation of number of different
				types of method parameters in every method divided by a multiplication
				of number of different method parameter types in whole class and
				number of methods. \\\hline
				cbm &coupling between methods &  total number of new/redefined methods
				to which all the inherited methods are coupled\\\hline
				cbo & coupling between objects & increased when the methods of one
				class access services of another.\\\hline
				ce & efferent couplings & how many other classes is used by the
				specific class. \\\hline
				dam & data access & ratio of the number of private (protected)
				attributes to the total number of attributes\\\hline
				dit & depth of inheritance tree &\\\hline
				ic & inheritance coupling &  number of parent classes to which a given
				class is coupled (includes counts of methods and variables inherited)
				\\\hline
				lcom & lack of cohesion in methods &number of pairs of methods that do
				not share a reference to an case variable.\\\hline
				locm3 & another lack of cohesion measure & if $m,a$ are  the number of
				$methods,attributes$
				in a class number and $\mu(a)$  is the number of methods accessing an
				attribute, 
				then
				$lcom3=((\frac{1}{a} \sum, j^a \mu(a, j)) - m)/ (1-m)$.
				\\\hline
				loc & lines of code &\\\hline
				max\, cc & maximum McCabe & maximum McCabe's cyclomatic complexity seen
				in class\\\hline
				mfa & functional abstraction & number of methods inherited by a class
				plus number of methods accessible by member methods of the
				class\\\hline
				moa &  aggregation &  count of the number of data declarations (class
				fields) whose types are user defined classes\\\hline
				noc &  number of children &\\\hline
				npm & number of public methods & \\\hline
				rfc & response for a class &number of  methods invoked in response to
				a message to the object.\\\hline
				wmc & weighted methods per class &\\\hline
				\rowcolor{lightgray}
				nDefects & raw defect counts & Numeric: number of defects found in post-release bug-tracking systems.\\
				\rowcolor{lightgray}
				isDefective & defects present? & Boolean: if {\em nDefects} $>0$ then {\em true} else {\em false}
			\end{tabular}
		}
	\end{center}
	\caption{OO code metrics used for all studies in this paper.
	   Last lines, shown in \textcolor{gray}{gray}, denote the dependent variables.}\label{fig:ck}
\end{figure*}
	
	A  blank cell in \fig{smells}
	indicates where   other work omits
	one of the bad smells in column one.
	Note that most of the cells are blank, and that the studies omit the 
	majority of the Fowler bad smells.
	SonarQube has no detectors for many of the column one bad smells.
	Also, nearly half the Yamashita list of bad smells
	does not appear in column 1. The eight numbers
	in the  Yamashita'13 column show the rankings for the bad smells
	that overlap with Fowler and Kerievsky; Yamashita also discussed other 
	smells not covered in Fowler'99~\cite{fowler99}.

	Two of the studies in \fig{smells} offers some comments on the relative 
	importance
	of the different bad smells. Three of the bad smells listed in the top 
	half 
	of the Yamashita'13 rankings also score very high in the developer survey. 
	Those three were {\em duplicated code, large class},
	and {\em long method}.
	Note that this agreement also means that the
	Yamashita'13 study and the developer survey
	believe
	that very few  code smells are   high priority issues
	requiring code reorganization.
	
	In summary, just because one developer strongly believes in the importance 
	of a bad smell, it does not mean that belief transfers to other developers 
	or 
	projects.
	Developers can be clever, but their thinking can also be distorted
	by cognitive biases.
	Hence, as shown in \fig{smells}, developers, text books, and tools
	can disagree on which bad smells are important.
	Special tools are needed to assess their beliefs, for example, their 
	beliefs in
	bad smells.

	{\color{steel}
		\section{Why Use Defect Data?}
		\label{sect:dataset}
		To assess our planning methods, we opted to use data gathered by 
		Jureczko et al. for object-oriented JAVA systems~\cite{jureczko10}. 
		The 
		``Jureczko'' data records the number of known defects for each class 
		using a post-release bug tracking system. The classes are described in 
		terms of nearly two dozen metrics such as number of children (noc), 
		lines of code (loc), etc. For details on the Jureczko data, see 
		\fig{ck}. The nature of collected data and  its relevance to defect 
		prediction is discussed in greater detail by Madeyski \& 
		Jureczko~\cite{madeyski15}.
		
		A sample set of values from a data set (ant 1.3) is shown 
		in 
		\fig{example}. Each instance notes a class under 
		consideration, 20 static code metric values, and 2 columns counting the 
		defects in the code. 
		
		The term ``defect'' in our work always refers to 
		\textit{defective 
		classes}. The defects themselves are represented in two ways: 
		
		\be
		
		\item Raw defect counts (denoted as \texttt{nDefects}): This refers 
		total number of defects present in a given class. This is an integer 
		value, such that, $nDefects~\in~\{0,1,2,\ldots\}$. This~representation 
		of defects is used by all the \textit{primary change oracles}.
		\item Boolean defects (denoted as \texttt{isDefective}): This is a 
		Boolean representation of defects. It is \texttt{TRUE} if 
		$\texttt{nDefects}>~0$ otherwise it is 
		\texttt{FALSE}.~This~representation of defects is used by 
		the \textit{secondary verification oracle}.
		
		\ee
		
		We use defects to operationalize smell definitions 
		following the 
		findings of several researchers. Li \& Shatnawi~\cite{li07} 
		investigated the relationship between the bad smells and module defect 
		probability. Their study found that, in the context of the 
		post-release 
		system evolution process, bad smells were positively associated with 
		the defect probability in the three error-severity levels. They also 
		reported that Shotgun Surgery, God Class, and God Method smells are 
		associated with higher levels of defects. Olbrich et 
		al.~\cite{olbrich2010all} show the impact that God and Brain Class 
		smells have on code directly influences defects in systems.
		
		In a more recent study, Hall et al.~\cite{Hall2014} further corroborate 
		the claim that smells indicate defect-prone code in several 
		circumstances. Their evaluation of smell detection performance shows 
		that it is difficult to define and quantify smell definitions, either 
		for automatic or manual smell detection. Generally, agreement levels 
		on 
		what code contains a smell are poor between tools, between tools and 
		humans, and even between humans. This in general leads to poor 
		performance of tools that detect code smells. More importantly, they 
		note that arbitrary refactoring is unlikely to significantly reduce 
		fault-proneness and in some cases may increase fault-proneness. Our 
		findings also show that smells have different effects on different 
		systems many of which cannot be quantified with just code smell 
		metrics. These findings lead us to conclusion that using static code 
		metrics and associated defects would best support code reorganization.}

		\begin{figure}[tp!]
\centering
\resizebox{\linewidth}{!}{
\label{my-label}
\begin{tabular}{c|l|l|l|c|l|c}
& & \multicolumn{3}{c|}{Metrics} & \multicolumn{2}{c}{Class} \\ \hline
Version                & \multicolumn{1}{c|}{Module Name}                             & wmc                    & dit                    & \ldots & nDefects & isDefective \\ \hline
1.3                    & org.apache.tools.ant.taskdefs.ExecuteOn & 11                     & 4                      & \ldots & 0        & FALSE       \\ \hline
1.3                    & org.apache.tools.ant.DefaultLogger      & 14                     & 1                      & \ldots & 2        & TRUE        \\ \hline
\multicolumn{1}{c|}{\ldots} & \multicolumn{1}{c|}{\ldots}                  & \multicolumn{1}{c|}{\ldots} & \multicolumn{1}{c|}{\ldots} & \ldots & 0        & FALSE       \\ \hline
1.3                    & org.apache.tools.ant.taskdefs.Cvs       & 12                     & 3                      & \ldots & 0        & FALSE       \\ \hline
1.3                    & org.apache.tools.ant.taskdefs.Copyfile  & 6                      & 3                      & \ldots & 1        & TRUE       
\end{tabular}}
\caption{A sample of ant 1.3}
\label{fig:example}
\end{figure}
	\section{Learning Bad Smell Thresholds}\label{sect:bst}
	
	Having made the case for automated, evidence-based support for assessing 
	bad smells,
	this section reviews different ways for building those tools (one of those 
	tools,
	XTREE, will be our recommended {\em primary change oracle}).
	Later in this paper, we describe  a {\em secondary verification oracle}
	that checks the effectiveness of the proposed changes.
	
	The SE literature offers two ways of learning bad smell thresholds.
	One approach relies on
	{\em outlier statistics}~\cite{erni96,bender99}. This approach
	has been used   by Shatnawi~\cite{Shatnawi10}, Alves et al.~\cite{Alves2010}
	and Hermans et al.~\cite{hermans15}.
	Another approach is
	based on {\em cluster deltas} that we developed
	for   Centroid Deltas~\cite{me12c} and
	is used here for XTREE.
	These two approaches are discussed below.
	
	\subsection{Outlier Statistics}
	
	The outlier approach assumes that unusually large measurements indicate 
	risk-prone code.
	Hence, they generate one bad smell threshold for any metric
	with such an ``unusually large'' measurement.
	The literature lists several ways to define ``unusually large''.

	\subsubsection{Enri \& Lewerentz}
	Given classes described with the  code metrics of \fig{ck},
	Enri and Lewerentz~\cite{erni96} found the   mean $\mu$ and the standard 
	deviation $\sigma$
	of each
	code metrics. Their definition of problematic outlier was any code
	metric with a measurement greater than $\mu+\sigma$.
	Shatnawi and Alves et al.~\cite{Shatnawi10,Alves2010} depreciate
	using $\mu+\sigma$ since it does not consider the fault-proneness of 
	classes when the thresholds are computed. Also, the approach lacks  
	empirical 
	verification.
	
	\subsubsection{ Shatnawi}\label{sect:shatnawi}
	Shatnawi~\cite{Shatnawi10}'s preferred alternative to $\mu+\sigma$
	is to extend the use VARL (Value of Acceptable Risk Level) which was 
	initially proposed by 
	Bender~\cite{bender99} for his epidemiology studies.  This approach uses two
	constants ($p_0$ and $p_1$) to compute the thresholds which, following 
	Shatnawi's guidance, we set to
	$p_0=p_1=0.05$.

	\begin{figure*}[t!]
~\hrule~
\begin{minipage}{.53\linewidth}
\small
 \begin{tabular}{p{0.95\linewidth}} 
 On the right-hand-side is a tree
 generated by iterative dichomization. 
 This tree can be read like a nested if-then-else statement; e.g.
 \begin{itemize}
     \item Lines 3 and 8 show two branches for lines of code (denoted here as {\em `\$loc}) below 698 and above 698.
     \item Any line with a colon ":" character shows  a  leaf  of  this  nesting.   For  example,  if  some  new  code module is passed down this tree and falls to the line marked in \textcolor{orange}{{\bf orange}},  the colon on that line indicates a prediction that this module has a 100\% chance of being defective. 
     \end{itemize}
Using this tree,
 XTREE looks for a nearby branch that has a lower chance of being defective. Finding the \textcolor{green}{{ green}} desired  branch,  XTREE reports a bad smell threshold for that module that is the delta between the \textcolor{orange}{{\bf orange}}  current branch  and   \textcolor{green}{{ green}} designed branch.
 In this case, that threshold relates to:
 \begin{itemize}
     \item Lines of code and comments ({\em lcom}) 
     \item The cohesion between classes ({\em cam}) which measures similarity of parameter lists to assess the relatedness amongst class   methods.
     \end{itemize}\\ 
 \end{tabular}
 \end{minipage}~~~~
 \begin{minipage}{.45\linewidth}
\includegraphics[width=\linewidth]{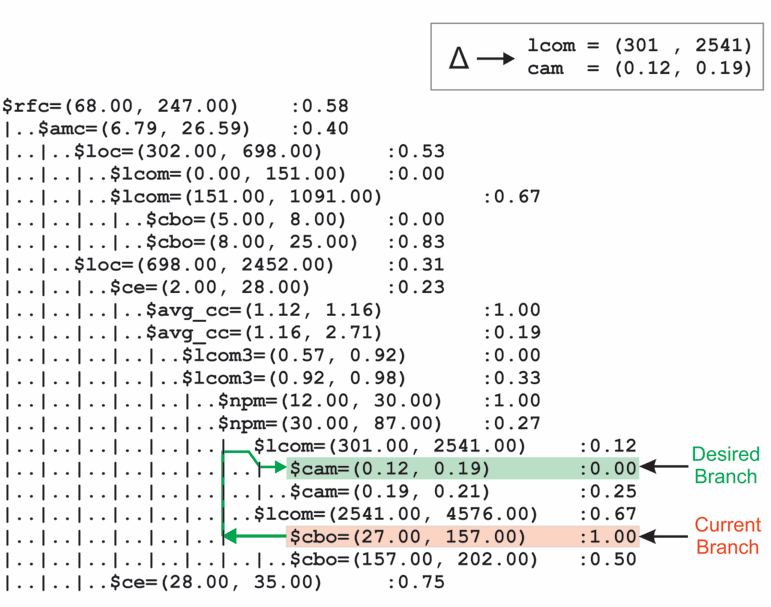}
\end{minipage}
~\hrule~
 \caption{A brief tutorial on XTREE.} \label{fig:xtree_samp}
\end{figure*}
	VARL encodes the defect count
	for each class as 0 (no defects known in class) or 1 (defects known in 
	class).
	Univariate binary logistic regression is applied to learn three 
	coefficients:
	$\alpha$ is the intercept constant;
	$\beta$ is the coefficient for maximizing log-likelihood;
	and $p_0$
	measures   how well this  model predicts for  defects. A univariate 
	logistic regression was conducted comparing metrics to defect counts. Any 
	code metric with $p>0.05$ is ignored as being a poor defect predictor. 
	Thresholds are then learned from the surviving metrics $M_c$ using
	the risk equation proposed by Bender:
	$$ \mathit{bad\; smell\; if\;} M_c > VARL$$
	Where,
	\begin{equation}
		VARL = p^{-1}(p_0) =  \frac{1}{\beta }\left( {\log \left( 
		{\frac{{{p_1}}}{{1 - {p_1}}}} \right) - \alpha } \right)
	\end{equation}

	\subsubsection{ Alves et al.}
	Alves et al.~\cite{Alves2010} propose another approach
	that  uses the underlying statistical 
	distribution and scale of the metrics.
	Metric values are weighted according to the source lines of 
	code (SLOC) of the class. All the weighted metrics are then normalized by 
	the sum of all weights for the system.
	The normalized metric values are ordered in an ascending fashion (this is 
	equivalent to computing a density function, in which the x-axis represents 
	the weight ratio (0-100\%), and the y-axis the metric scale).
	Alves et al. then select a percentage value (they suggest 70\%) which 
	represents the ``normal'' values for metrics. The metric threshold, then, 
	is the metric value for which 70\% of the classes fall below. The 
	intuition  is that the worst code has outliers beyond 70\% of the normal 
	code measurements i.e., they state that the risk of there existing a defect 
	is moderate to high when the threshold value of 70\% is exceeded.

	Hermans et al.~\cite{hermans15} used this approach in their  2015 paper on
	exploring bad smells. We explore the correlation between the code metrics 
	and the defect counts with a univariate logistic regression and  reject 
	code metrics that are poor predictors of defects (i.e.   those  with $p > 
	0.05$).
	
	\subsubsection{Discussion of Outlier Methods}\label{sect:disc}
	The advantage of the outlier-based
	approaches is that they are simple to implement, but the approaches have   
	two  major disadvantages.
	First, they are {\em verbose}. A threshold can be calculated for every 
	metric -- so, which one should the developers focus on changing? Without a 
	means for prioritizing the  thresholds and metrics against one another, 
	developers may have numerous or conflicting recommendations on what to 
	improve. Second, the outlier approaches suffers from {\em conjunctive 
	fallacy}  discussed in the introduction. That is, while	they propose 
	thresholds for many code metrics
	individually, they make no comment on what minimal metrics need to be 
	changed at the same time (or whether or not those changes lead to 
	minimization or maximization of static code measures).

	\subsection{Cluster Deltas}
	
	Cluster deltas are a general framework
	for learning {\em conjunctions} of changes
	that need to be applied at the same time.
	This approach works as follows:
	\begin{itemize}
		\item Cluster the data.
		\item Find
		neighboring clusters $C_+,C_-$ (where $C_+$ has more examples of 
		defective
		modules than $C_-$);
		\item Compute the  delta   in code metrics between the clusters using 
		\mbox{$\Delta = C_- - C_+ = \left\{\delta|\delta\in C_-, \delta \notin 
		C_+\right\}$}, i.e.
		{\em towards} the cluster with lower defects;
		\item The set $\Delta$ are changes needed in defective modules of $C_+$ 
		to
		make them more like the less-defective modules of $C_-$
	\end{itemize}
	Note that $\Delta$ is a conjunction of  recommendations.
	Since it is computed
	from neighboring clusters, the examples contain similar distributions and 
	$\Delta$ respects the naturally occurring constraints in the data. For 
	example,
	given a bad smell pertaining to large methods,   $\Delta$   will not  
	suggest lowering lines of code
	without also increasing a coupling measure.
	Cluster deltas are used in CD~\cite{me12c} and XTREE.

	\subsubsection{CD}\label{sect:cdcd}
	Borges and Menzies first proposed CD centroid deltas to
	generate {\em conjunctions} of code metrics
	that need to be changed at the same time
	in order to reduce defects~\cite{me12c}.
	CD uses the WHERE clustering algorithm developed by the
	authors for a prior application~\cite{me12d}.
	Each cluster was then replaced by its centroid
	and $\Delta$ was calculated directly from the difference
	between code metric values between one centroid
	and its nearest neighbor.

	One drawback with CD is that it is {\em verbose}
	since
	CD   recommended changes to all code
	metrics with different values in those two centroids.
	This makes it hard to use CD to   critique and prune away bad smells. 
	Further, CD will be shown not to be as effective
	in proposing changes to reduce defects as XTREE.
	
	\subsubsection{XTREE: Mining Project History for Defect-Proneness 
	Attributes}
	\label{sect:xtreextree}
	XTREE  is a cluster delta algorithm
	that avoids the problem of verbose $\Delta$s.
	XTREE is our {\em primary change oracle} that makes recommendations
	of what changes should be made to code modules.
	Instead of reasoning over cluster centroids,
	XTREE utilizes a decision tree learning approach
	to find the fewest differences between clusters of examples.

	XTREE uses a multi-interval discretizer based on an iterative 
	dichotomization scheme, first proposed by Fayyad and Irani~\cite{fi}. This 
	method converts the values for each code metric into a small number of 
	nominal ranges. It works as follows:
	\begin{itemize}
		\item A code metric is split into $r$ ($r=2$) ranges, each range is of
		size $n_r$ and is associated with a set of defect counts $x_r$ with 
		standard deviation
		$\sigma_r$.
		\item The best split for that range is the one that minimizes the 
		expected value of the
		defect variance, after the split; i.e. $\sum_r\frac{n_r}{n}\sigma_x$ 
		(where $n=\sum_r n_r$).
		\item This discretizer then recurses on  split to find 
		other splits in a recursive fashion. As suggested by Fayyad and Irani, 
		minimum description length (MDL) is used as a termination criterion for 
		the recursive partitioning.
	\end{itemize}
	
	When discretization finishes, each code metric $M$ has a
	final expected value $M_v$ for the defect standard deviation
	across all the discretized ranges of that metric.
	Iterative dichomization sorts the metrics by $M_v$
	to find the code metric that best splits the data i.e., the code metric 
	with smallest $M_v$.
	
	A decision tree is then constructed on the discretized metrics. The metric 
	that generated the best split forms the root of the tree with its discrete 
	ranges acting as the nodes.
	
	When all the metrics are arranged this way, the process is very similar to 
	a hierarchical clustering algorithm that groups together code modules with 
	similar defect counts and some shared ranges of code metrics.
	For our purposes, we score each cluster found in this way according
	to the percent of classes with known defects. For example, the last line 
	of 
	\fig{xtree_samp} shows a tree leaf with 75\%
	defective modules.
	
	\fig{xtree_samp} offers
	a small example of how XTREE builds
	$\Delta$ by comparing branches that lead to leaf clusters
	with different defect percentages. In this example, assume a project with 
	a 
	table of code metrics data describing its classes in the form of 
	\fig{example}. 
	After code inspections and running test cases or operational
	tests, each such class is augmented with a defect count.
	Iterative dichomization takes that table of data and,
	generates the tree of \fig{xtree_samp}.
	
	Once the tree is built, a class with code metric data is passed into the 
	tree and evaluated down the tree to a leaf node (see the 
	\textcolor{orange}{{\bf orange}} line in \fig{xtree_samp}).
	XTREE then looks for a nearby leaf node with a lower defect
	count (see the \textcolor{green}{{green}} line in \fig{xtree_samp}). For 
	that evaluated class, XTREE proposes bad smell
	thresholds that are the differences between
	\textcolor{green}{{green}} and \textcolor{orange}{{\bf orange}}.

	\subsubsection{XTREE: Recommending Code Reorganizations}
		\label{sect:xtreextree2}
	Using the training data construct a decision tree as suggested above.
	
	Next, for each test code module,
	find $C_+$ as follows: take each test, run it down the decision tree to 
	find a leaf in the decision tree that best matches the test case.
	After that,	find $C_-$ as follows:
	\begin{itemize}
		\item Starting at the  $C_+$ leaf (level N), ascend $lvl\in 
		\{N,N-1,N-2, 
		...\}$ tree 
		levels;
		\item Identify {\em sibling} leaves; i.e. leaf clusters that can be 
		reached from level $lvl$ that are not same as {\em current $C_+$};
		\item Find the {\em better} siblings; i.e. those with defect proneness 
		50\% or less than that of $C_+$ (e.g., if defect-proneness of $C_+$ is 
		70\%, find the nearest sibling with defect proneness $\leq$ 35\%). If 
		none found, then repeat for $lvl$--$=1$. Also, return \texttt{nil} if 
		the new $lvl$ is above the root.
		\item  Set $C_-$ to the  {\em closest} better sibling where distance is 
		measured between the mean centroids of that sibling and {\em $C_+$}
	\end{itemize}
	Now find $\Delta = C_- - C_+$  by reflecting
	on the set difference between  conditions in the decision tree branching 
	from $C_+$ to $C_-$. To find that delta,
	for discrete attributes, delta is the value of the {\em desired}; for 
	numerics  expressed as ranges, the delta could be any value that lies 
	between $\left( LOW, HIGH\right]$ in that range.
	
	Note that XTREE's recommendation does not exhaustively search the tree for 
	the change that reduces defect proneness the most, but rather finds the 
	nearest sibling. This is by design. This design allows XTREE to a) 
	provide 
	recommendations quickly, and b) to recommend changes to a small number of 
	attributes.
	
	\section{Setup}
	\label{sect:setup}
	The previous section proposed numerous methods for detecting bad smells 
	that need
	to be resolved. This section offers a way to evaluate them as follows:
	\bi
	\item Use each framework discussed in \tion{bst} as a {\em primary change 
	oracle} to recommend how 
	code should be changed in our test data set (Section~\ref{sect:tesd}).
	\item Apply those changes. (This is emulated by changing the code metrics 
	in order that all the $\Delta$'s are addressed.)\footnote{Note: This was 
	done by selecting a random number from the $\left(LOW,  HIGH\right]$ 
	boundaries of the recommended delta(s). The random number indicates what 
	may happen if a developer were to make changes to their code to comply with 
	the recommendations.}
	\item Run a {\em secondary verification oracle} to assess the 
	defect-proneness of the changed code.
	\item Sort the change oracles on how well they reduce defects as judged by 
	the verification oracle.
	\ei
	Using this, we can address the research questions discussed in the 
	introduction.
	
	{\bf RQ1: Effectiveness: } Which of the methods defined in Section 3 is the 
	best change oracle for identifying what and how code modules should be 
	changed?
	We will assume that developers
	update and reorganize their code until the bad smell thresholds are not 
	violated.
	This code reorganization will start with some {\em initial} code
	base that is changed to a {\em new} code base. 
	
	\rahul{
	For example, assume that a log 
	history of defects has shown that modules 
	with \mbox{{\em loc $>100$}} have more defects (per class) than smaller 
	modules and a
	code module has 500 lines of code. The action is to reduce the size of that 
	module; we reason
	optimistically that we can change that code metric
	to 100.  Using the secondary verification oracle,  we then predict the
	number of defects in {\em new}.}
	
	We compare $d_+$, the number of defects in the \textit{initial} code base 
	to $d_-$, the number of defects in the \textit{new} code base.
	We evaluate the performance of XTREE, CD, Shatnawi, and Alves change 
	oracles. The best change oracle is the one that maximizes
	\begin{equation}\label{eq:diff}
		\mathit{improvement} = 100* \left(1 - \frac{ d_- }{ d_+}\right)
	\end{equation}

	{\bf RQ2: Succinctness: } Which of the Section 3 methods recommended 
	changes to the fewest code attributes?
	To answer this question, we will report the frequency at which different 
	attributes are selected in repeated runs of our oracles.
	
	{\bf RQ3: Stopping: }    How effective is XTREE at offering   ``stopping 
	points'' (i.e. clear guidance on what {\em not} to do)?
	To answer this point, we will report how often XTREE's recommendations {\em 
	omit} a code attribute. Note that the {\em more often} XTREE omits an 
	attribute, the more likely it is {\em not} to endorse addressing a bad 
	smell based on the omitted attributes.
	
	{\bf RQ4: Stability:} Across different projects, how variable are the 
	changes recommended by XTREE?   To answer this
	question, we conduct a large scale ``what if'' study that reports all the 
	possible recommendations XTREE might make. We then count how often 
	attributes are {\em not} found in the recommendations arising from this 
	``what if'' study.
	
	{\bf RQ5: Conjunctive Fallacy:} Is  it  always  useful  to  apply \eq{df}; 
	i.e.   make  code  better  by  reducing  the  values  of multiple code 
	attributes? To answer this question, we will look at the {\em direction of 
	change} seen in the {\bf RQ4} study; i.e.
	how often XTREE recommends decreasing or increasing a static code attribute.
	
	\subsection{Test Data}\label{sect:tesd}
	
	To explore these research questions, we used data from Jureczko et al.'s 
	collection of object-oriented Java systems~\cite{jureczko10}. To access 
	that data, go to   git.io/vGYxc.
	The Jureczko data records the number of known defects for each class using 
	a post-release bug tracking system. The classes are described in terms of 
	nearly two dozen metrics included in the Chidamber and Kemerer metric 
	suite, such as number of children (noc), lines of code (loc), etc. For 
	details on the Jureczko code
	metrics, see  \fig{ck} and corresponding text in \tion{dataset}. For 
	details on the versions of that data that were used for training and 
	testing purposed see the left-hand-side columns of \fig{j}.

	\subsection{Building the Secondary Verification Oracle}
	\label{sect:eval}
	
	As mentioned in the introduction, our proposed framework has two oracles:
	a primary change oracle (XTREE) and the secondary verification oracle 
	described in this section.
	
	\rahul{
	It can be difficult  to judge the  effects of removing bad 
	smells.
	Code that is reorganized cannot be assessed just by a rerun of the test
	suite for three reasons: (1) Test suites may not 
	be 
	designed to identify code smells, they only report pass/fail based on 
	whether or not a specific module runs. It is entirely possible that a test 
	case may pass for code that contains one or more code smells; (2) While 
	smells are certainly symptomatic of design flaws, not all smells cause the 
	system to fail in manner suitable for identification by test cases; and (3) 
	It make take a significant amount of effort to write new test cases that 
	identify bad smells, especially for relatively stable software systems. 
	Additionally, since we 
	can not be sure if reorganizations do/don't change the system behavior, for 
	instance cases where a reorganization effort involves only refactoring, it 
	becomes difficult to assert when a test case will pass or fail.}

		\begin{figure*}[hbtp!]
	\small
	\begin{center}
		\begin{minipage}{.46\linewidth}
			\begin{tabular}{r@{~}|l@{~}|r@{~}|l@{~}|r@{~}|r@{~}|} \cline{2-6}

				& \multicolumn{5}{c|}{ Data set  properties}\\ 
			 
				& \multicolumn{2}{c|}{training}   & \multicolumn{3}{c|}{testing}      \\ \cline{2-6}
				data set      & versions           & cases & versions     & cases    & \% defective             \\ \hline
				jedit    & 3.2, 4.0, 4.1, 4.2 & 1257      & 4.3          & 492          & 2 \\
				ivy      & 1.1, 1.4           & 352       & 2.0          & 352          & 11 \\
				camel    & 1.0, 1.2, 1.4      & 1819      & 1.6          & 965          & 19 \\
				ant      & 1.3, 1.4, 1.5, 1.6 & 947       & 1.7          & 745          & 22 \\
				synapse  & 1.0, 1.1           & 379       & 1.2          & 256          & 34 \\
				velocity & 1.4, 1.5           & 410       & 1.6          & 229          & 34 \\
				lucene   & 2.0, 2.2           & 442       & 2.4          & 340          & 59 \\
				poi      & 1.5, 2, 2.5        & 936       & 3.0          & 442          & 64 \\
			 xerces   & 1.0, 1.2, 1.3      & 1055      & 1.4          & 588          & 74  \\ 
			 log4j    & 1.0, 1.1           & 244       & 1.2          & 205          & 92   \\
			 xalan    & 2.4, 2.5, 2.6      & 2411      & 2.7          & 909          & 99  \\\hline

			\end{tabular}\end{minipage}~~~~~~\begin{minipage}{.4\linewidth}
			\begin{tabular}{|rrr|rrr|rr|l} \cline{1-8}
			 
				\multicolumn{8}{|c|}{  Results from learning}\\
			 
				\multicolumn{3}{|c|}{untuned} & \multicolumn{3}{c|}{tuned} & \multicolumn{2}{c|}{change}\\
				\cline{1-8}
				
				pd & pf & good? & pd & pf & good? & pd & pf\\\cline{1-8}
				\rowcolor{celadon}55 & 29 &   & 64 & 29 & y & 9 & 0&$\star$\\
				\rowcolor{celadon}	65 & 35 & y & 65 & 28 & y & 0 & -7&$\star$\\
				49 & 31 &   & 56 & 37 &   & 5 & 6\\
				\rowcolor{celadon}	49 & 13 & y & 63 & 16 & y & 14 & 3&$\star$\\
				45 & 19 &   & 47 & 15 &   & 2 & -4\\
				78 & 60 &   & 76 & 60 &   & -2 & 0\\
				56 & 25 &   & 60 & 25 & y & 4 & 0\\
				\rowcolor{celadon}	56 & 31 &   & 60 & 10 & y & 4 & -21&$\star$\\
			\rowcolor{lavenderpink}	30 & 31 &   & 40 & 29 &   & 10 & -2&$\times$\\
				\rowcolor{lavenderpink}32 & 6 &   & 30 & 6 &   & -2 & 0&$\times$\\
				\rowcolor{lavenderpink}38 & 9 &   & 47 & 9 &   & 9 & 0&$\times$\\
				\hline 
			\end{tabular}
			
		\end{minipage}
	\end{center}    
	
	\caption{Training and test {\em data set properties} for  Jureczko data ,
		sorted by \% defective examples.
		On the right-hand-side, we show the {\em results from learning}.
		Data is usable if it has a recall of 60\% or more and false alarm of 30\% or less (and note that, after tuning, there are more usable data sets than before). Results  	\colorbox{celadon}{ marked with ``$\star$''} show large improvements in performance, after tuning
		(lower {\em pf} or higher {\em pd}).
		Data in  the  \colorbox{lavenderpink}{three bottom rows}, marked with ``$\times$'', are  performing
		poorly-- that data so many defective examples  that it  is hard for
		our learners to distinguish between classes.
	}\label{fig:j}
\end{figure*}

	To resolve this problem, SE researchers such as
	Cheng et al.~\cite{Cheng10}, O'Keefe et al.~\cite{OKeeffe08,OKeeffe07},
	Moghadam~\cite{Moghadam2011} and Mkaouer et al.~\cite{Mkaouer14}
	use a {\em secondary verification oracle} that is learned separately
	from the primary oracle. The verification oracles assesses
	how defective the code is before and after some
	code reorganization.
	For their second oracle,
	Cheng, O'Keefe, Moghadam and  Mkaouer et al. use the QMOOD hierarchical
	quality model~\cite{Bansiya02}.
	A shortcoming of QMOOD
	is that quality models learned from other projects
	may perform poorly when applied to new projects~\cite{me12d}.
	Hence, for this study, we  eschew
	older quality models like QMOOD. Instead, we use
	Random Forests~\cite{Breiman2001} to learn defect predictors
	from OO code metrics.
	Unlike QMOOD, the predictors
	are specific to the project.
	% 		, and the measurements can be reconstructed and the predictors 
	%rebuilt for other projects.
	
	Random Forests are a decision tree learning method but
	instead of building one tree, hundreds are built using
	randomly selected subsets of the data. The final predictions
	come from averaging the predictions over all the trees.
	Recent studies endorsed the use
	of  Random Forests for  defect prediction~\cite{lessmann}.
	
	\fig{j} shows  our results with Random Forests and
	the Jureczko data. The goal is to build a verification oracle based on 
	Random Forest that accurately distinguishes between defective and 
	non-defective files based on code metrics. Given $V$ released versions, we 
	test on version $V$ and train on the available data from $V-1$ earlier 
	releases. 
	
	The \colorbox{lavenderpink}{three bottom rows}  are marked 
	with $\times$: these contain predominately defective classes (two-thirds, 
	or more).  It is hard to build a model that distinguishes non-defective 
	files from defective files in these data sets due to the high percentage of 
	defective file examples.
	
	We use Boolean classes in the  Jureczko data to identify the presence or 
	absence of defects: \texttt{True} if defects \textgreater 0; \texttt{False} 
	if defects = 0. The quality of the predictor is measured using (a) the  
	probability of detection \textit{pd} (i.e., recall):  the percent of faulty 
	classes in the test data detected by the {\em predictor}; and (b) the  
	probability of false alarm \textit{pf} (i.e., false positives): the percent 
	of non-fault classes that are {\em predicted} to be defective.
	
	\rahul{
	\subsubsection{Impact of Tuning}
	\label{sect:tune}
	
	The ``untuned'' columns of \fig{j}
	show a preliminary study using Random Forest with its ``off-the-shelf'' 
	tuning of 100 trees per forest.
	The forests were built from training data and applied to test data
	not seen during training.  In this
	study, we called a data set ``usable'' if   Random Forest was able to 
	classify the instances with a performance threshold of $\mathit{pd}\ge 60 
	\wedge \mathit{pf} \le 30$\% (determined from standard results in other 
	publications~\cite{me07b}). We found that no  data set satisfy
	this criteria.
	
	To salvage Random Forest we first applied the SMOTE algorithm to improve 
	the 
	performance of the classifier by handling class imbalance in the data sets. 
	Pelayo 
	and Dick~\cite{pelayo07} report that defect prediction is improved by 
	SMOTE~\cite{chawla2002smote}; i.e. an over-sampling of minority-class 
	examples and an under-sampling of majority-class examples. 
	
	In a recent paper, Fu et al.~\cite{fu:ase15} reported that parameter 
	tuning with differential evolution~\cite{storn97} can quickly explore the 
	tuning options of Random Forest to find better settings for the size of the 
	forest, the termination criteria for tree generation, and other parameters. 
	In a setting similar to theirs, we implemented a multiobjective 
	differential evolution to tune Random Forests for each of the above 
	datasets.
	
	Setting goals for tuning can be a very difficult task. Fu et 
	al. warn that choosing goals must be undertaken with caution. Opting to 
	optimize individual goals such as recall/precision of a learner can have 
	undesirable outcomes. For instance, by tuning for recall we can achieve 
	near 100\% recall -- but at the cost of a near 100\% false alarms. On the 
	other hand, when we tune for false alarms, we can achieve near zero percent 
	false alarm rates by effectively turning off the detector (so the 
	recall falls to nearly zero). Therefore, in our work, we used used a 
	``multiobjective'' search to tune for both recall and false alarm at the 
	same time. A multiobjective search attempts to find settings that achieve a 
	\textit{trade-off} between the goals (in this case maximizing recall and 
	minimizing false alarm at the same time). 
	
	The effect of tuning and using SMOTE were remarkable.The rows 
	\colorbox{celadon}{marked with a $\star$} in \fig{j} show data 
	sets whose performance was improved by these techniques. For 
	example, in {\em poi}, the recall increased by 4\% while the false alarm 
	rate dropped by 21\%. In \textit{Ivy, Jedit, Lucene,} and \textit{Xalan} 
	there was a 
	significant improvement in one measure (recall or false alarm) with no 
	deterioration in the other measure. Finally, in \textit{Poi} and 
	\textit{Ant} there was a significantly large improvement in one metric with 
	a very small deterioration in the other. }

	However, as expected, we found that some 
	datasets (xerces, 
	xalan, log4j, 
	...) were not responsive to our tuning efforts. Since, we could not salvage 
	all the data 
	sets, we eliminated these data sets for which we could not build 
	an adequately performing Random Forest classifier with $\mathit{pd}\ge 60 
	\wedge \mathit{pf} \le 30$\%. Thus, our analysis uses the {\em jedit, ivy, 
	ant, lucene} and {\em poi} data sets for evaluating recommended changes.
	
	We note that SMOTE-ing and
	parameter tunings were applied to the training data only and not to the 
	test data.
	
	\subsection{Statistics}

	We use 40 repeated runs for each code reorganization recommendation 
	framework 
	for each of the five data sets (we use 40 since that is  more than the 30 
	samples  needed to satisfy the central limit theorem). Each code 
	organization framework is trained on versions $1...N-1$ of $N$ available 
	versions in each data set.
	Each run collects the improvement scores defined in \eq{diff}: the 
	reduction in the percentage of defective classes as identified by the 
	verification oracle after applying the recommended code metric changes.
	
	We use multiple runs with different random number seeds since two of our 
	methods use some random choices: CD uses the  stochastic WHERE clustering 
	algorithm~\cite{me12d}
	while XTREE non-deterministically picks thresholds randomly from
	the high and low boundary of a range.
	Hence, to compare all
	four methods, we must run the analysis many times.

	\begin{figure}[tbp]
%{\scriptsize \textbf{Ant}\\[0.1cm]}

{\scriptsize \hspace{3.5cm}\underline{Observed Improvements (from \eq{diff})}}\vspace{2mm}

{\scriptsize \textbf{Ant}~~~~~~~~ \begin{tabular}{{l@{~~~~}l@{~~~~}|r@{~~~~}r@{}c@{~~~}r}}
\arrayrulecolor{lightgray}
\rowcolor{lightgray}\textbf{Rank} & \textbf{Treatment} & \textbf{Median} & \textbf{IQR~~~} & \\
  1 &         XTREE &    56   &  21  & \quart{54}{25}{65}{1} \\
\hline  2 &        Alves &    32   &  17  & \quart{28}{20}{37}{1} \\
\hline  3 &     Shatnawi &    15   &  4.2 & \quart{15}{6}{18}{1} \\
  3 &           CD &    12   &  0  & \quart{15}{0}{15}{6} \\
\hline \end{tabular}}\\

%{\scriptsize \textbf{Lucene}\\[0.1cm]}
%{\scriptsize \textbf{Poi}\\[0.1cm]}
{\scriptsize 

\textbf{Poi}~~~~~~~~ \begin{tabular}{{l@{~~~~}l@{~~~~}|r@{~~~~}r@{~~}c@{}r}}
\arrayrulecolor{lightgray}
\rowcolor{lightgray}\textbf{Rank} & \textbf{Treatment} & \textbf{Median} & \textbf{IQR~~~} & \\
        1 &         XTREE &    20   &  16  & \quart{39}{40}{51}{2} \\
\hline  2 &        Alves &    14   &  16  & \quart{21}{41}{37}{2} \\
\hline  3 &           CD &    11   &  0  & \quart{23}{0}{23}{7} \\
        3 &     Shatnawi &    8   &  1  & \quart{19}{5}{21}{2} \\
\hline \end{tabular}}\\

{\scriptsize \textbf{Lucene}~ \begin{tabular}{{l@{~~~~}l@{~~~~}|r@{~~~~}r@{~~}c@{}r}}
\arrayrulecolor{lightgray}
\rowcolor{lightgray}\textbf{Rank} & \textbf{Treatment} & \textbf{Median} & \textbf{IQR~~~} & \\
        1 &         XTREE &    16   &  6  & \quart{50}{29}{71}{4} \\
        1 &     Shatnawi  &    15   &  2  & \quart{63}{10}{67}{4} \\
\hline  2 &           CD  &    13   &  0  & \quart{55}{0}{55}{6} \\
\hline  3 &        Alves  &    9   &  4  & \quart{33}{19}{42}{4} \\
\hline \end{tabular}}\\

%{\scriptsize \textbf{Ivy}\\[0.1cm]}
{\scriptsize \textbf{Ivy}~~~~~~~~ \begin{tabular}{{l@{~~~~}l@{~~~~}|r@{~~~~}r@{~~}c@{}r}}
\arrayrulecolor{lightgray}
\rowcolor{lightgray}\textbf{Rank} & \textbf{Treatment} & \textbf{Median} & \textbf{IQR~~~} & \\
        1 &        Alves &    67   &  20  & \quart{58}{21}{71}{1} \\
\hline  2 &         XTREE &    52   &  22  & \quart{42}{24}{55}{1} \\
\hline  3 &           CD &    35   &  0  & \quart{27}{0}{27}{2} \\
\hline  4 &     Shatnawi &    20   &  7  & \quart{18}{8}{21}{1} \\
\hline \end{tabular}}\\

%{\scriptsize \textbf{Jedit}\\[0.1cm]}

{\scriptsize  \textbf{Jedit}~~~~~ \begin{tabular}{{l@{~~~}l@{~~~~}|r@{~~~~}r@{~~}c@{}r}}
\arrayrulecolor{lightgray}
\rowcolor{lightgray}\textbf{Rank} & \textbf{Treatment} & \textbf{Median} & \textbf{IQR~~~} & \\
  1 &        Alves &    36   &  7  & \quart{60}{10}{66}{1} \\
  1 &        XTREE &    36   &  0  & \quart{66}{0}{66}{2} \\
  1 &     Shatnawi &    36   &  9  & \quart{53}{13}{66}{1} \\
  1 &          CD &    36   &  0  & \quart{66}{0}{66}{2} \\
\hline \end{tabular}}\\
\caption{Results for {\bf RQ1} from the
Jureczko   data sets.  Results from 40 repeats.
Values come from \eq{diff}.
Values near 0
imply no improvement,
{\em Larger} median values are {\em better}. 
Note that XTREE and Alves are usually best and CD and Shatnami
are usually worse.}
\label{fig:jur2}
\end{figure}

	To rank these 40 numbers collected from CD, XTREE, Shatnawi, and Alves et 
	al., we use the Scott-Knott test recommended by Mittas and 
	Angelis~\cite{mittas13}.
	Scott-Knott is a top-down clustering approach used to rank different
	treatments. If that clustering finds an interesting division of the data, 
	then
	some statistical test is applied to the two divisions to check if they
	are statistically significant different. If so, Scott-Knott recurses
	into both halves.
	
	To  apply Scott-Knott,
	we
	sorted a list of  $l=40$ values of \eq{diff} values found in  $ls=4$ 
	different methods.
	Then, we split $l$ into sub-lists $m,n$ in order to maximize the expected 
	value of differences in the observed performances before and after 
	divisions. E.g. for lists $l,m,n$ of size $ls,ms,ns$ where $l=m\cup n$: 
	\[E(\Delta)=\frac{ms}{ls}abs(m.\mu - l.\mu)^2 + \frac{ns}{ls}abs(n.\mu - 
	l.\mu)^2\]
	We then apply a apply a statistical hypothesis test $H$ to check
	if $m,n$ are significantly different  (in our case, the conjunction of A12 
	and bootstrapping). If so, Scott-Knott recurses on the splits. In other 
	words, we divide the data if \textit{both} bootstrap sampling and effect 
	size test agree that a division is statistically significant (with a 
	confidence of 99\%) and not a small effect ($A12 \ge 0.6$).
	For a justification of the use of non-parametric bootstrapping, see Efron 
	\& Tibshirani~\cite[p220-223]{efron93}. For a justification of the use of 
	effect size tests see Shepperd\&MacDonell~\cite{shepperd12a}; 
	Kampenes~\cite{kampenes07}; and Kocaguenli et 
	al.~\cite{Kocaguneli2013:ep}. 
	These researchers warn that even if a hypothesis test declares two 
	populations to be ``significantly'' different, then that result is 
	misleading if the ``effect size'' is very small. Hence, to assess the 
	performance differences we first must rule out small effects using A12, a 
	test   recently endorsed by Arcuri and Briand at ICSE'11~\cite{arcuri11}.
	
	The Scott-Knott  results are presented in the form of line diagrams like 
	those shown on the right-hand-side of \fig{jur2}.
	The black dot shows the median \eq{diff} values and the horizontal lines 
	stretches
	from the 25\textsuperscript{th} percentile to the 75\textsuperscript{th} 
	percentile (the inter-quartile range, IQR).
	As an example of how to read this table, consider the {\em Ant}
	results. Those rows are  sorted on the median values of each framework. 
	Note 
	that all the methods have \eq{diff}~\textgreater~$0\%$; i.e. all these 
	methods reduced the expected value of the performance score while XTREE 
	achieved the greatest reduction (of 56\% from the original value).
	These results table has a  left-hand-side  {\bf Rank} column, computed 
	using the
	Scott-Knott test described above. In the {\em Ant}
	results, XTREE is ranked the best, while CD is  ranked   worst.
	
% \usepackage{multirow}
\newcommand{\ZZ}{.}
\begin{figure*}
\renewcommand{\baselinestretch}{0.8} 
\scriptsize  
\centering

\label{my-label}
                  
\begin{tabular}{c|rrrr|rrrr|rrrr|rrrr|rrrr}
\multicolumn{1}{c}{\cellcolor[HTML]{EFEFEF}{\color[HTML]{000000} }} & \multicolumn{4}{c}{\cellcolor[HTML]{EFEFEF}{\color[HTML]{000000} Ant}} & \multicolumn{4}{c}{\cellcolor[HTML]{EFEFEF}{\color[HTML]{000000} Ivy}} & \multicolumn{4}{c}{\cellcolor[HTML]{EFEFEF}{\color[HTML]{000000} Lucene}} & \multicolumn{4}{c}{\cellcolor[HTML]{EFEFEF}{\color[HTML]{000000} Jedit}} & \multicolumn{4}{c}{\cellcolor[HTML]{EFEFEF}{\color[HTML]{000000} Poi}} \\
\multicolumn{1}{c@{}}{\multirow{-2}{*}{\cellcolor[HTML]{EFEFEF}{\color[HTML]{000000} Features}}} & \multicolumn{1}{c@{}}{\cellcolor[HTML]{EFEFEF}{\color[HTML]{000000} XTREE}} & \multicolumn{1}{c@{}}{\cellcolor[HTML]{EFEFEF}{\color[HTML]{000000} CD}} & \multicolumn{1}{c@{}}{\cellcolor[HTML]{EFEFEF}{\color[HTML]{000000} Alves}} & \multicolumn{1}{c@{}}{\cellcolor[HTML]{EFEFEF}{\color[HTML]{000000} Shatn}} & \multicolumn{1}{c@{}}{\cellcolor[HTML]{EFEFEF}{\color[HTML]{000000} XTREE}} & \multicolumn{1}{c@{}}{\cellcolor[HTML]{EFEFEF}{\color[HTML]{000000} CD}} & \multicolumn{1}{c@{}}{\cellcolor[HTML]{EFEFEF}{\color[HTML]{000000} Alves}} & \multicolumn{1}{c@{}}{\cellcolor[HTML]{EFEFEF}{\color[HTML]{000000} Shatn}} & \multicolumn{1}{c@{}}{\cellcolor[HTML]{EFEFEF}{\color[HTML]{000000} XTREE}} & \multicolumn{1}{c@{}}{\cellcolor[HTML]{EFEFEF}{\color[HTML]{000000} CD}} & \multicolumn{1}{c@{}}{\cellcolor[HTML]{EFEFEF}{\color[HTML]{000000} Alves}} & \multicolumn{1}{c@{}}{\cellcolor[HTML]{EFEFEF}{\color[HTML]{000000} Shatn}} & \multicolumn{1}{c@{}}{\cellcolor[HTML]{EFEFEF}{\color[HTML]{000000} XTREE}} & \multicolumn{1}{c@{}}{\cellcolor[HTML]{EFEFEF}{\color[HTML]{000000} CD}} & \multicolumn{1}{c@{}}{\cellcolor[HTML]{EFEFEF}{\color[HTML]{000000} Alves}} & \multicolumn{1}{c@{}}{\cellcolor[HTML]{EFEFEF}{\color[HTML]{000000} Shatn}} & \multicolumn{1}{c@{}}{\cellcolor[HTML]{EFEFEF}{\color[HTML]{000000} XTREE}} & \multicolumn{1}{c@{}}{\cellcolor[HTML]{EFEFEF}{\color[HTML]{000000} CD}} & \multicolumn{1}{c@{}}{\cellcolor[HTML]{EFEFEF}{\color[HTML]{000000} Alves}} & \multicolumn{1}{c@{}}{\cellcolor[HTML]{EFEFEF}{\color[HTML]{000000} Shatn}} \\
wmc & \ZZ & 92 & 100 & 100 & 18 & 95 & 100 & 100 & 89 & 95 & 100 & \ZZ &\ZZ& 63 & \ZZ &\ZZ& \ZZ & 100 & 100 &\ZZ\\
dit & \ZZ & 77 & 100 & \ZZ & \ZZ & 87 & 100 & \ZZ & \ZZ & 80 & 100 & \ZZ &\ZZ& 72 & 100 & 100 & \ZZ & 46 & 100 &\ZZ\\
noc & \ZZ & 20 & 100 & \ZZ & \ZZ & \ZZ & 100 & \ZZ & \ZZ & 26 & \ZZ &\ZZ& \ZZ &\ZZ& \ZZ &\ZZ& \ZZ &\ZZ& \ZZ &\ZZ\\
cbo & 88 & 99 & 100 & 100 & 91 & 100 & 100 & 100 & 60 & 94 & 100 & 100 & \ZZ & 100 & 100 & \ZZ & 1 & 74 & 100 &\ZZ\\
rfc & 100 & 100 & 100 & \ZZ & 8 & 95 & 100 & \ZZ & 10 & 83 & 100 & \ZZ & 100 & 100 & 100 & 100 & 100 & 95 & 100 &\ZZ\\
lcom & \ZZ & 98 & 100 & 100 & 15 & 100 & 100 & 100 & \ZZ & 94 & 100 & \ZZ &\ZZ& 100 & 100 & \ZZ &\ZZ& 100 & 100 & 100 \\
ca & \ZZ & 93 & 100 & \ZZ & 7 & 95 & 100 & \ZZ & 40 & 89 & 100 & \ZZ &\ZZ& 63 & 100 & 100 & \ZZ & 74 & 100 &\ZZ\\
ce & 5 & 100 & 100 & \ZZ & \ZZ & 97 & 100 & \ZZ &\ZZ& 90 & 100 & \ZZ &\ZZ& 100 & 100 & 100 & \ZZ & 64 & 100 &\ZZ\\
npm & \ZZ & 88 & 100 & \ZZ & 8 & 97 & 100 & \ZZ &\ZZ& 93 & 100 & 100 & \ZZ & 100 & 100 & \ZZ &\ZZ& 100 & 100 &\ZZ\\
lcom3 & \ZZ & 90 & 100 & \ZZ & 7 & 95 & 100 & \ZZ & 13 & 79 & 100 & 100 & \ZZ & 63 & 100 & 100 & \ZZ & 92 & 100 & 100 \\
loc & 100 & 99 & 100 & 100 & 97 & 97 & 100 & 100 & 60 & 100 & 100 & 100 & \ZZ & 100 & \ZZ & 100 & 100 & 100 & 100 & 100 \\
dam & \ZZ & 21 & 100 & \ZZ & \ZZ & 22 & 100 & \ZZ &\ZZ& 55 & 100 & \ZZ &\ZZ& 45 & 100 & 100 & \ZZ & 73 & 100 &\ZZ\\
moa & \ZZ & 67 & 100 & \ZZ & \ZZ & 82 & 100 & \ZZ &\ZZ& 60 & 100 & 100 & \ZZ & 54 & 100 & 100 & \ZZ & 58 & 100 &\ZZ\\
mfa & 5 & 93 & 100 & \ZZ & \ZZ & 90 & 100 & \ZZ & 5 & 80 & 100 & \ZZ &\ZZ& 72 & 100 & \ZZ &\ZZ& 72 & 100 &\ZZ\\
cam & \ZZ & 99 & 100 & 100 & 84 & 100 & 100 & 100 & 10 & 94 & 100 & \ZZ &\ZZ& 100 & 100 & \ZZ &\ZZ& 98 & 100 & 100 \\
ic & \ZZ & 52 & 100 & 100 & \ZZ & 70 & \ZZ &\ZZ& \ZZ & 68 & 100 & \ZZ &\ZZ& 36 & 100 & \ZZ &\ZZ& 43 & 100 & 100 \\
cbm & \ZZ & 59 & 100 & \ZZ & \ZZ & 85 & \ZZ &\ZZ& \ZZ & 71 & 100 & \ZZ &\ZZ& 36 & 100 & 100 & \ZZ & 67 & 100 &\ZZ\\
amc & \ZZ & 99 & \ZZ & \ZZ & \ZZ & 95 & 100 & \ZZ & 30 & 100 & \ZZ & \ZZ & \ZZ & 100 & 100 & 100 & \ZZ & 97 & \ZZ &\ZZ\\
max cc & \ZZ & 87 & 100 & 100 & \ZZ & 85 & 100 & 100 & \ZZ & 71 & \ZZ &\ZZ& \ZZ & 45 & 100 & \ZZ &\ZZ& 63 & 100 &\ZZ\\
avg cc & 12 & 99 & 100 & 100 & \ZZ & 95 & 100 & 100 & 13 & 98 & \ZZ &\ZZ& \ZZ & 100 & 100 & 100 & \ZZ & 92 & 100 & \ZZ\\\hline
\end{tabular}
\caption{Results for {\bf RQ2}.
Percentage counts of  how often an approach recommends changing a code metric
(in 40 runs). ``100'' means that this code metric
was always recommended. Cells marked with ``.'' indicate  0\%. For the Shatnawi and Alves et al.
columns,  metrics score 0\% if they always fail the $p \le 0.05$ test of \tion{shatnawi}.
For CD, cells are blank when two centroids have the same value for the same code
metrics. For XTREE, cells are blanks when they do not appear in the delta
between branches.  Note
that XTREE mentions specific code metrics
far fewer times than other methods.}\label{fig:counts}
\end{figure*}
	\section{Results}
	\label{sect:results}
	
	\begin{figure*}[!t]
		\centering
		\includegraphics[width=\linewidth]{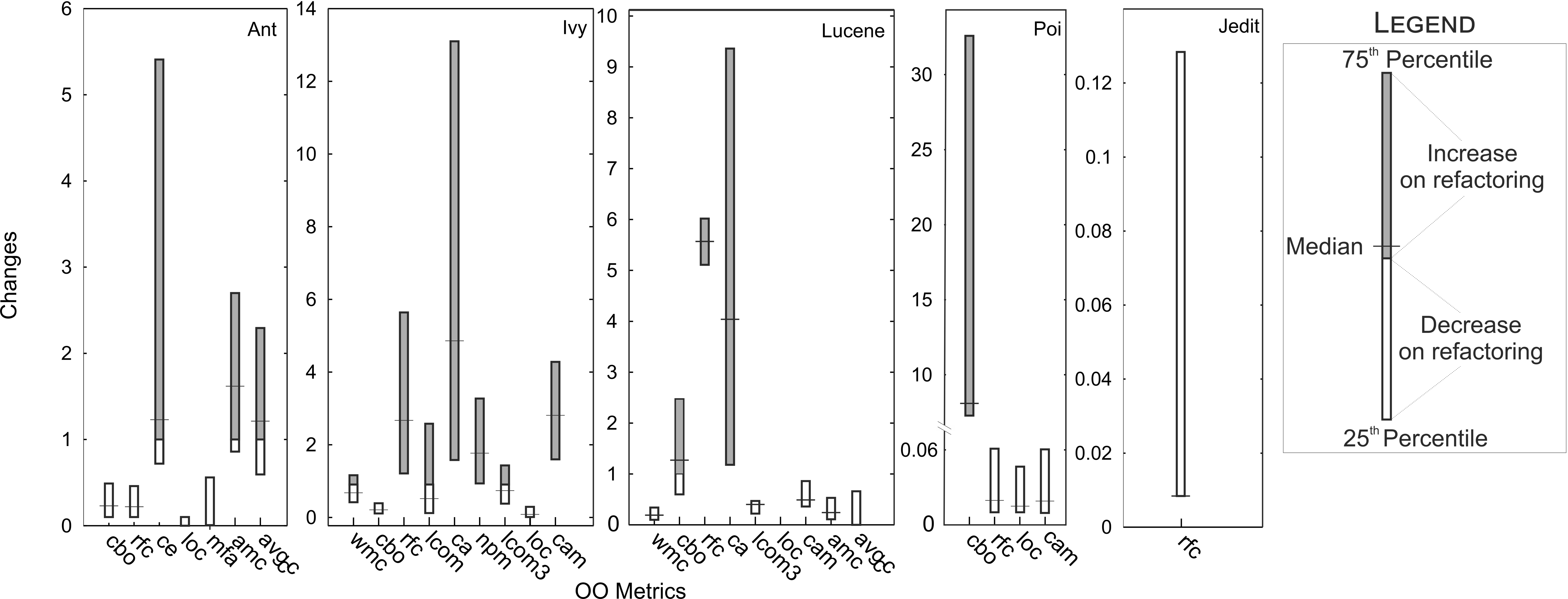}
		\caption{Results  from XTREE.
			While \fig{counts} are the {\em number} of times a code metric was 
			changed,
			this  figure shows the {\em magnitude} each code metric was 
			changed. Each vertical bar
			marks the 27,50,75\textsuperscript{th} percentile change seen in 40 
			repeats.
			All numbers are ratios of initial to final values.
			All bar regions marked in gray show {\em increases}.
			The interesting feature of these results are that many
			of the changes proposed by XTREE require {\em increases}
			(this puzzling observation is explained in the text).}
		\label{fig:changes}
	\end{figure*}

	\subsection{RQ1: Effectiveness}
	
	{\em Which of the methods defined in Section 3 is the best change oracle 
	for identifying what and how code modules should be changed? }
	
	\fig{jur2} shows the comparison results.
	Two data sets are very responsive to defect reduction suggestions:
	Ant and Ivy (both of which show best case improvements over 50\%).
	The  expected value of defects
	is changed less in Jedit. This data set's results are
	surprisingly uniform; i.e.   all methods
	find the same ways to reduce the expected number of
	defects.
	
	\fig{changes} enables us to explain the uniformity
	of the results seen with Jedit in \fig{jur2}.
	Observe how in \fig{changes} the only change ever
	found is a reduction to {\em rfc}. Clearly, in this
	data set, there is very little that can be usefully changed.
	
	Two data sets are not very responsive to defect reduction:
	Poi and Lucene. The reason for this can be see in \fig{j}:
	both these data sets contain more than 50\% defective modules.
	In that space, all our  methods lack a large sample of
	defect-free examples.
	
	Also consider the relative
	rank of the different approaches,
	CD and Shatnawi usually  perform comparatively worse while  XTREE gets top 
	ranked position the most
	number of times. That said, Alves sometimes beats XTREE (see Ivy)
	while sometimes it ties (see Jedit).
	
	In summary, our {\bf Result1} is  that, of the change oracles studies here,
	XTREE is the best oracle on how to change code modules in order to reduce 
	defects in our data sets.
	
% \usepackage{graphicx}
\begin{figure*}
\centering
\resizebox{\textwidth}{!}{%
\begin{tabular}{l|c|c|c|c|c|c|c|c|c|c|c|c|c|c|c|c|c|c|c|c}
         & wmc & dit & noc & cbo & rfc & lcom & ca & ce & npm & lcom3 & loc & dam & moa & mfa & cam & ic & cbm & amc & max\_cc & avg\_cc \bigstrut\\ \hline
Ant      &  & & & $-$ &$-$&  & &$+$& & &$-$& & &$-$& & &  &$+$&     &$+$\bigstrut\\ \hline

Ivy      &$-$& & &$-$&$+$&$-$&$+$& &$-$&$-$&$-$& & & &$+$& & & &     &      \bigstrut\\ \hline

Poi      & & & &$+$&$-$& & & & & &$-$& & & &$-$& & & & & \bigstrut\\ \hline

Lucene   &$-$& & &$+$&$+$& &$+$& & &$-$&$-$& & & &$-$& & &$-$& &$-$\bigstrut\\ \hline

Jedit    & & & & &$-$& & & & & & & & & & & & & & &     \bigstrut\\ 

\end{tabular}}

\caption{Direct of changes seen in  
a comparison of statistically significantly different static code attributes measures seen in the clusters found by XTREE. Each dataset contains 20 Static Code Metrics (for a description of each of these metrics, please refer to~\cite{me12d}). The rows contain the datasets, and the columns denote the metrics. A ``$+$'' symbol represents a recommendation that requires a significant statistical increase (with a p-value$\le$0.05), and likewise, a ``$-$'' represents a significant statistical decrease.} 
\label{fig:multcomp}
\end{figure*}
	
	\subsection{RQ2: Verbosity}
	
	{\em Which of the Section 3 methods recommended changes to the fewest code 
	attributes?}
	
\fig{counts} shows the frequency with which the methods
recommend changes to specific code metrics.
Note that XTREE proposes thresholds to
few code metrics compared to the other approaches.
	
	% COMMENT From LL: This is such a strange way of putting this lesson, and 
	%you haven't actually evaluated this.
	%
	
	Hence, our {\bf Result2} is that, of all the code change oracles 
	studied here, XTREE recommended far fewer changes to static code 
	measures. Note only that,   combining  \fig{jur2} with \fig{counts}, 
	we   see that
	even though XTREE proposes changes to far fewer code metrics, those few
	code metrics are usually just as effective (or
	more effective) than the multiple
	thresholds
	proposed by CD, Shatnawi or Alves.  That is, XTREE proposes
	{\em fewer} and better thresholds than the other approaches.

	\subsection{RQ3: Stopping}
	
	{\em  How effective is XTREE at offering   ``stopping points'' (i.e. clear 
	guidance on what not to do)?}
	
	The {\bf RQ2} results showed that XTREE's recommendations are small in a 
	{\em relative sense}; i.e. they are
	relatively smaller than the other methods studied here.
	Note also that, XTREE's recommendations are small in an {\em absolute 
	sense}.
	Consider the frequency of changes in \fig{counts}. There are very few 
	values for XTREE that are over 33\% (in our sense this translates to at 
	least a third of our repeated runs where XTREE mentioned a code 
	attribute). 
	For Ant, Ivy, Lucene, Jedit, and Poi, those frequencies
	are  \mbox{3, 3, 3, 4, 1, 2} respectively (out of twenty). This means 
	that, 
	usually, XTREE omits references to \mbox{17,17,17,16,19,18} static code 
	attributes (out of 20). Any code reorganization based on a bad smell 
	detector that uses  these omitted code attributes could hence  be stopped.
	
	Hence our {\bf Result3} is that, in a any  project,  XTREE's  recommended  
	changes affect only one to four of the 20 static code attributes.  Any bad 
	smell defined in terms of the remaining 19 to 16 code attributes (i.e. 
	most 
	of them) would be deprecated.

	\subsection{RQ4: Stability}
	
	{\em Across different projects, how variable are the changes recommended by 
	our best change oracle? }

	\fig{counts} counted how {\em often} XTREE's recommendations mentioned a 
	static code attribute.
	\fig{changes}, on the other hand, shows the {\em direction} of XTREE's 
	recommended change:
	\bi
	\item Gray bars show  an  {\em increase} to a static code measure;
	\item White bars shows a   {\em decrease} to a static code measure;
	\item Bars that are all white or all gray indicate that in our 40 repeated 
	runs, XTREE recommended changing an attribute the same way, all the time.
	\item Bars that are mixtures of white and gray mean that, sometimes, XTREE 
	makes different recommendations about how to change a static
	code attribute.
	\ei
	Based on \fig{changes}, we see {\bf Result4} states that the direction of 
	change recommended  XTREE is  very stable repeated runs of the program  
	(evidence:
	the bars are mostly the same color).
	
	\fig{changes} also comments on the inter-relationships between static code 
	attributes. Note that while some measures in
	\fig{changes} are decreased, many are increased.
	For example, consider the {\em Poi} results
	from \fig{counts} that recommends decreasing {\em loc}
	but making large increases to {\em cbo} (coupling between
	objects). Here, XTREE is advising us to break up
	large classes class by into services
	in other classes. Note that such a code reorganization will, by
	definition, increase the coupling between objects.
	Note also that such increases  to reduce
	defects would never be proposed by the outlier methods
	of Shatnawi or Alves since their core assumption is that bad
	smells arise from unusually large code metrics.
	
	\subsection{RQ5: Conjunctive Fallacy}
	\label{sect:conjunct}
	{\em  Is  it  always  useful  to  apply \eq{df};  i.e.   make  code  
	better  by  reducing  the  values  of multiple code attributes?}
	
	\rahul{
	In  \fig{changes}, the bars are colored both white and gray; i.e. XTREE 
	recommends {\em decreasing} and {\em increasing}
	static attribute values. That is, always decreasing static code measures 
	(as suggested by \eq{df}) is {\em not} recommended by XTREE. This comments 
	on the interconnectedness of metrics and it is important for a few reasons:
	
	\be
	\item When developers follow suggestions offered based on static code 
	metrics, they are often required to reduce several metrics at once. For 
	example, in \fig{counts} we notice all methods except XTREE recommend 
	changes to \textit{every} metric. This is not a practical approach and it 
	renders these suggestions useless.
	\item When XTREE recommends changes, it respects the interconnectedness of  
	metrics. If we were to reduce LOC, other metrics to do with coupling (cbo, 
	ca, ce, ...) would have to be changed as well. XTREE is aware of this and 
	suggests changes to 
	these metrics in conjunction with LOC. Additionally, it offers a direction 
	of change (increase or decrease). 
	\ee
	
	Now when programmers follow the recommendations of XTREE instead for 
	reorganization, they do so fully aware of the impact that change can have 
	on other metrics. For instance, a programmer who chooses to reduce LOC is 
	now aware that a consequence would be increased coupling. She/he can 
	continue to make this change because similar changes with lower 
	LOC and higher coupling have historically been known to result in fewer 
	defects.

	}

	To explore this point further, we conducted the following ``what-if'' 
	study. Once XTREE builds its trees with its leaf clusters then:
	\bi
	\item Gather all clusters $C_0$ for $C_1$ such that the percent of defects 
	in $C_0$ 
	is greater than $C_1$ \ldots
	\item For all attribute measures, identify ones that have a statistically 
	significantly different distribution between each $C_0$ and $C_1 $...
	\item Report the direction of change ($+$ indicates an \textit{increase} in 
	values 
	and $-$ indicates a \textit{decrease})
	\ei

	{\color{steel}
	Note that the changes found in this way are somewhat more general than the 
	results presented in~\fig{jur2}. Those results were limited to comments on 
	the test set given to the program given that the tree is constructed using 
	the training set. 
	
	Performing a ``what-if'' analysis using the three steps presented above 
	allow us to reflect on {\em all possible changes} for any dataset (not 
	limited to only one instance from the test set). This is significant 
	because it comments on some of the commonly held notions regarding static 
	code attributes. Figure ~\ref{fig:multcomp} which shows the results of this 
	``what if'' analysis can better help understand this. Consider as an 
	example the metric loc, as might be expected, the recommendation is to
	\textit{always} reduce lines of code (loc). But for the other attributes 
	like afferent and efferent coupling, contrary to popularly recommendations,
	XTREE in fact suggests to increase those values or leave them unchanged, 
	but \textit{not} decrease. Similarly, coupling between objects (cbo) needs 
	to be decreased in Ant and Ivy but increased in Poi and Lucene. Finally, 
	many metrics such as depth of inheritance tree (dit), are best left 
	unchanged in all cases. 
	
	This is quite significant for practitioners 
	attempting to perform code reorganization to similar projects without a 
	historical log. They can use these findings as guide to: (1) deprecate 
	changes that show no demonstrable benefits; and (2) not limit themselves to 
	reducing values all the metrics because an increase can sometimes be more 
	beneficial.

	Hence, {\bf Result5} is that while XTREE always recommends reducing loc,
	it also   often recommends increasing the values of other static code 
	attributes.}

	\section{Limitations and Future Work}
	\label{sect:future}
	\rahul{
					
		XTREE offers to assist reorganization efforts in an intelligent 
		fashion, it suffers from two key limitations: (1) They use supervised 
		learning to learn about the domain, therefore they need a sufficiently 
		large dataset so as to be ``trained''. \fig{j} shows the dataset used 
		in this study all have at least two prior releases; (2) Ability to 
		build a reliable secondary verification oracle. We used the current 
		state of the art in build our verification oracle (Tuned and SMOTE-ed 
		Random Forest).

        In this work, both the primary change oracle and the secondary 
        verification oracle 
        used OO metrics in conjunction with the number of associated defects to 
        operationalize code smells. We note this goal can very easily be 
        replaced by other quantifiable objectives. For example, there is much 
        research on \textit{technical debts} in recent years.
        The notion of technical debt reflects the extra work arising
        from developers making pragmatic short-term decisions
        that make it harder, in the long-term, to maintain the software.
        A study by 
        Ernst et al.~\cite{ernst15} showed that practitioners agree on the 
        importance of the issue but existing tools are not currently helpful in 
        managing the debts. A systematic literature survey conducted by Li et 
        al.~\cite{li15td} report that dedicated technical debt management (TDM) 
        tools are needed for managing various types of technical debts. 
        Recently Alves et al.~\cite{alves2016} 
        conducted an 
        extensive mapping study on the identification and management of 
        technical debt. They identified several types of debts over the past 
        decade and found that bad smells of the sort discussed in this paper 
        are very well known indicators of design debt. In fact, they happen to 
        be few of most 
        frequently referenced kinds of debts. 
        XTREE 
        could be adapted for this purpose by supporting decisions if and when a 
        technical debt item should be paid off -- providing that there existed
        a working ``secondary oracle'' (of the kind we define in our introduction)
        that can recognize quick-and-dirty sections of code.}

		Our research shows that it is potentially na\"{\i}ve to explore 
		thresholds in static code attributes in isolation to each other. 
		This work clearly demonstrates how changing one necessitates changing 
		other associated metrics. So, for future work, we recommend that 
		researchers and practitioners look for tools that recommend changes to 
		sets 
		of code changes. When exploring candidate technologies, apart from 
		XTREE, 
		researchers may potentially use: (1) Association rule learning; (2) 
		Thresholds generated across synthetic dimensions (eg. PCA); and (3) 
		Techniques that cluster data and look for deltas between them. (Note: 
		we 
		offer XTREE as a possible example of this point).
		
		Lastly, we also
		plan on extending our work by exploring scalable 
		solutions 
		to achieve similar results in much larger datasets. After this, we 
		shall 
		look at applications beyond that of measuring static 
		code attributes. For example, as an initial attempt, we have been 
		looking 
		at sentiment analysis in stack overflow exchanges to learn dialog 
		pattern 
		that most select for relevant entries.

	\section{Reliability and Validity of Conclusions}
	\label{sect:valid}

	The results of this paper are biased by our choice of code reorganization
	goal (reducing defects) and our choice of measures collected from software
	project (OO measures such as depth of inheritance, number of child classes,
	etc). That said, it should be possible extend the methods of this paper to 
	other
	kinds of goals (e.g. maintainability, reliability, security, or the 
	knowledge sharing
	measures) and other kinds of
	inputs (e.g. the process measures favored by Rahman,
	Devanbu et al.~\cite{Rahman2013})
	
	\subsection{Reliability}
	Reliability refers to the consistency of the results obtained
	from the research. It has at least two components: internal
	and external reliability.
	
	Internal reliability checks if an independent researcher
	reanalyzing the data would come to the same conclusion.
	To assist other researchers exploring this point, we offer a full 
	replication package for this study at
	https://github.com/ai-se/XTREE\_IST.
	
	External reliability assesses how well independent researchers
	could reproduce the study. To increase external
	reliability, this paper has taken care to clearly define our
	algorithms. Also, all the data used in this work is available
	online.
	
	For the researcher wishing to reproduce our work to other kinds of goals, 
	we offer the following advice:
	
	\bi
	\item Find a data source for the other measures of interest;
	\item Implement another secondary verification oracle that can assess 
	maintainability, reliability, security, technical debt, etc;
	\item Implement a better primary verification oracle that can do ``better'' 
	than XTREE at finding changes (where ``better'' is defined in terms
	of the opinions of the verification oracle). 
	\ei

	\subsection{Validity}
	
	This paper is a case study that studied the effects of  limiting 
	unnecessary code reorganization on some data sets. This section discusses 
	limitations of such case studies. In this context, validity refers to the 
	extent to which a piece of research actually
	investigates what the researcher purports to investigate.
	Validity has at least two components: internal and
	external validity.

	Based on the case study presented above,
	as well as the discussion in \tion{prelim},
	we believe that bad smell indicators (e.g. \mbox{{\em loc}$>$ 100})
	have limited external validity beyond the projects from which they are 
	derived.
	While specific models are externally valid,
	there may still be general methods like XTREE for finding the good local 
	models.
	
	Our definition of bad smells is limited to those represented by OO code 
	metrics (a premise often used in related work).
	XTREE, Shatnawi, Alves et al. can  only comment
	on bad smells   expressed as code metrics
	in the historical log of a project.
	
	If developers want to justify their code reorganizations
	via bad smells expressed in other terminology,
	then the  analysis of this paper must:
	\begin{itemize}
		\item Either wait till
		data about those new
		terms has been collected.
		\item Or, apply cutting edge transfer learning
		methods~\cite{Nam15,Jing15, krishna16} to map data from other projects
		into the current one.
	\end{itemize}
	Note that the transfer learning approach would
	be highly experimental and require more study
	before it can be safely recommended.
		
	Sampling bias threatens any data mining analysis; i.e., what matters
	there may not be true here. For example, the data sets used here comes 
	from 
	Jureczko et al. and any biases in their selection procedures
	threaten the validity of these results.
	That said,
	the best we can do is define our methods and publicize our data and code so 
	that other researchers can
	try to repeat our results and, perhaps, point out a previously unknown bias
	in our analysis. Hopefully, other researchers will emulate our methods in
	order to repeat, refute, or improve our results.

	\section{Conclusions}
	\label{sect:conclusions}
	How to discourage useless code reorganizations?
	We say:
        \begin{quote}{\em Ignore those changes
            not supported by the historical log of data from
	the current project.}\end{quote}
	When that data is not available (e.g. early
	in the project) then developers could use the general list of
	bad smells shown in \fig{smells}. However,
	our results
	show that  bad smell detectors are most
	effective when they are based
	on a small handful of code metrics (as done by XTREE).
	Hence, using all the bad smells of \fig{smells} may not be optimal.
	
	For our better guess at how to reduce defects by changing code attributes,
	see  Figure ~\ref{fig:multcomp}. But given the large variance if the 
	change 
	recommendations, we strongly advice teams to use
	XTREE on their data to find their own best local changes.
	
	XTREE improves on prior methods for generating bad smells:
	\begin{itemize}
		\item As described in \tion{prelim}, bad smells generated by humans may 
		not be applicable to the current project. On the other hand, XTREE can 
		automatically learn specific thresholds
		for bad smells for the current project.
		\item Prior methods used an old quality predictor (QMOOD) which we 
		replace with defect predictors learned via Random Forests
		from   current project data.

		\item XTREE's conjunctions proved to be arguably as effective
		as those of Alves (see \fig{jur2}) but far less verbose (see 
		\fig{counts}). Since XTREE {\em approves} of fewer changes
		it hence {\em disapproves} of most changes. This makes it a better 
		framework 
		for critiquing and rejecting
		many of the code reorganizations.
	\end{itemize}
	Finally, XTREE does not suffer from the conjunctive fallacy.
	Older methods, such as those proposed by  Shatnawi and Alves assumed
	that the best way to improve code is to remove outlier  values. This may 
	not work since when code is reorganized,
	{\em the functionality has to go somewhere}. Hence,  reducing the lines of 
	code in one module necessitates
	{\em increasing} the coupling that module
	to other parts of the code. In future, we recommend software team use bad 
	smell detectors that know what attribute measures
	need {\em decreasing} as well as {\em increasing}.
	
	% In summary, this paper offers a baseline result
	% that XTREE can be used as a {\em diagnostic tool}
	% for identifying which bad smells can be ignored.
	% Can it be used as a {\em prescriptive tool}
	% to recommend new refactorings? This would require
	
	% As discussed in {\em PROBLEM \#2}
	% (from the introduction), XTREE might propose
	% changes to code metrics that are
	% not feasible  (due to the local domain constraints
	% of the code). To repair this, XTREE needs to be extended
	% with architectural knowledge of the code it is studying.
	% Elsewhere, we have had some   success
	% with reasoning about such architectural knowledge
	% (expressed as software product lines~\cite{sayyad13a,sayyad13b}). In 
	%future
	% work, we will explore extending XTREE
	% with
	% architectural knowledge such that it can
	% become a tool which prescribes
	% feasible refactorings.

	\section*{Acknowledgements}
	The work is partially funded by NSF  awards \#1506586 and \#1302169.

	\balance
	\bibliographystyle{unsrt}
	\bibliography{References}

\end{document}